\documentclass[traditabstract]{aa}
\usepackage{graphicx,natbib}

\newcommand{\xspec}{{\sc xspec}}
\newcommand{\pexmon}{{\sc pexmon}}
\newcommand{\compps}{{\sc compps}}
\newcommand{\nthcomp}{{\sc nthcomp}}
\newcommand{\comptt}{{\sc comptt}}
\newcommand{\compst}{{\sc compst}}
\newcommand{\eqpair}{{\sc eqpair}}
\newcommand{\yhc}{$y_{hc}$}
\newcommand{\gwc}{$\Gamma_{wc}$}
\newcommand{\thc}{$kT_{hc}$}
\newcommand{\twc}{$kT_{wc}$}
\newcommand{\tahc}{$\tau_{hc}$}
\newcommand{\tawc}{$\tau_{wc}$}
\newcommand{\tbbhc}{$T_{bb,hc}$}
\newcommand{\tbbwc}{$T_{bb,wc}$}
\newcommand{\hot}{{\sc hot}}
\newcommand{\warm}{{\sc warm}}

\newcommand{\ergs}{erg\ s$^{-1}$}

\def\xmm{{\it XMM-Newton}}
\def\lta{ \lower .75ex\hbox{$\sim$} \llap{\raise .27ex \hbox{$<$}} }

\begin{document}
\date{Received .../Accepted ...}

\title{Multiwavelength campaign on Mrk 509\\ XII. Broad band spectral analysis.}
\author{P.-O. Petrucci\inst{1}
  \and S. Paltani\inst{2}
  \and J. Malzac\inst{3,4}
  \and J.S. Kaastra\inst{5,6}
  \and M. Cappi\inst{7}
  \and G. Ponti\inst{8}
  \and B. De Marco\inst{9}
  \and G.A. Kriss\inst{10,11}
  \and K.C. Steenbrugge\inst{12,13}
  \and S. Bianchi\inst{14}
  \and G. Branduardi-Raymont\inst{15}
  \and M. Mehdipour\inst{15}
  \and E. Costantini\inst{5}
  \and M. Dadina\inst{7}
  \and P. Lubi\'nski\inst{16}
  }
  
\institute{	   UJF-Grenoble 1 / CNRS-INSU, Institut de Plan\'etologie et d'Astrophysique
	   de Grenoble (IPAG) UMR 5274, Grenoble, F-38041, France
           \and
	   ISDC Data Centre for Astrophysics, Astronomical Observatory of the
	   University of Geneva, 16, ch. d'Ecogia, 1290 Versoix, Switzerland 
	   \and
	   Universit\'e de Toulouse, UPS-OMP, IRAP, Toulouse, France
	   \and
           CNRS, IRAP, 9 Av. colonel Roche, BP 44346, F-31028 Toulouse 
	   cedex 4, France
	   \and
	  SRON Netherlands Institute for Space Research, Sorbonnelaan 2,
           3584 CA Utrecht, the Netherlands 
	   \and
	   Sterrenkundig Instituut, Universiteit Utrecht, 
	   P.O. Box 80000, 3508 TA Utrecht, the Netherlands
	   \and
	   INAF-IASF Bologna, Via Gobetti 101, 40129 Bologna, Italy
	   \and
	   School of Physics and Astronomy, University of Southampton, 
	   Highfield, Southampton SO17 1BJ
	   \and
	   Centro de Astrobiolog\'{i}a (CSIC-INTA), Dep. de Astrof\'{i}sica; LAEFF, PO Box 78, E-28691, Villanueva de la Ca\~nada, Madrid, Spain 
	   \and
	   Space Telescope Science Institute, 3700 San Martin Drive, 
	   Baltimore, MD 21218, USA
	   \and
	   Department of Physics and Astronomy, The Johns Hopkins University,
	   Baltimore, MD 21218, USA
	   \and
           Instituto de Astronom\'ia, Universidad Cat\'olica del Norte, 
	   Avenida Angamos 0610, Casilla 1280, Antofagasta, Chile
	   \and
	   Department of Physics, University of Oxford, Keble Road, 
	   Oxford OX1 3RH, UK
	   \and 
	   Dipartimento di Fisica, Universit\`a degli Studi Roma Tre, 
	   via della Vasca Navale 84, 00146 Roma, Italy 
	   \and
	   Mullard Space Science Laboratory, University College London, 
	   Holmbury St. Mary, Dorking, Surrey, RH5 6NT, UK
	   \and
	   Centrum Astronomiczne im. M. Kopernika, 
	   Rabia\'nska 8, PL-87-100 Toru\'n, Poland
} 

%

\abstract{The origin of the different spectral components present in the high-energy (UV to X-rays/gamma-rays) spectra of Seyfert galaxies is still being debated a lot. One of the major limitations, in this respect, is the lack of really simultaneous broad-band observations that allow us to disentangle the behavior of each component and to better constrain their interconnections. The simultaneous UV to X-rays/gamma rays data obtained during the multiwavelength campaign on the bright Seyfert 1 Mrk 509  are used in this paper and tested against physically motivated broad band models.

Mrk 509 was observed by \xmm\ and INTEGRAL in October/November 2009, with one observation every four days for a total of ten observations. Each observation has been fitted with a realistic thermal Comptonization model for the continuum emission. Prompted by the correlation between the UV and soft X-ray flux, we used a thermal Comptonization component for the soft X-ray excess. We also included a warm absorber and a reflection component, as required by the precise studies previously done by our consortium.
The UV to X-ray/gamma-ray emission of Mrk 509 can be well fitted by these components. The presence of a relatively hard high-energy spectrum points to the existence of a hot ($kT \sim$ 100 keV), optically-thin ($\tau \sim$ 0.5) corona producing the primary continuum. In contrast, the soft X-ray component requires a warm ($kT \sim$ 1 keV), optically-thick ($\tau \sim$ 10-20) plasma. 

Estimates of the amplification ratio for this warm plasma support a configuration relatively close to the ``theoretical'' configuration of a slab corona  above a passive disk. An interesting consequence is the weak luminosity-dependence of its emission, which is a possible explanation of the roughly                      
constant spectral shape of the soft X-ray excess seen in AGNs. The temperature ($\sim$ 3 eV) and flux of the soft-photon field entering and cooling the warm plasma suggests that it covers the accretion disk down to a transition radius $R_{tr}$ of 10-20 $R_g$. This plasma could be the warm upper layer of the accretion disk.

In contrast, the hot corona has a more photon-starved geometry. The high temperature ($\sim$ 100 eV) of the soft-photon field entering and cooling it favors a localization of the hot corona in the inner flow. This soft-photon field could be part of the comptonized emission produced by the warm plasma. 
In this framework, the change in the geometry (i.e. $R_{tr}$) could explain most of the observed flux and spectral variability.}


\maketitle

\section{Introduction}
Since the discovery of a high-energy cut-off in the X-ray emission of Seyfert galaxies, first detected by
SIGMA and OSSE in NGC 4151 \citep{jou92,mai93} and then commonly observed in about 50\% of Seyfert galaxies \citep{zdz93,gon96,mat01,per02,del03,dad07,mol09}, the primary continuum of this class of AGN is commonly believed to be thermal Comptonization of soft seed photons produced by a `cold phase'',  presumably the accretion disk, on hot ($\sim$ 100 keV) thermal electrons (the ``hot phase'' also called corona). 
The different Compton scattering orders produced through this process eventually add together to  produce a roughly cut-off power-law shape (e.g. \citealt{sun80}), which agrees well with the high-energy ($> $2 keV) spectra of Seyfert galaxies.\\

The nature and origin of this thermal hot plasma are, however, still largely unknown, mainly owing to the lack of sensitive instruments at high energies. This hinders our ability to put tight constraints on the spectral shape and to disentangle the primary continuum from the reflection component. At lower energies ($<$ 2 keV), the presence of deep absorption features, the signatures of ionized plasma (the so-called warm absorber, WA hereafter) surrounding the central engine, and soft X-ray emission  in excess of the extrapolation of the high-energy continuum to these energies (the so-called soft X-ray excess) does not allow a direct view of the primary emission. Thus, we are generally left with the 2-10 keV band, which is far too narrow for a deep understanding of the origin of the high-energy broad-band emission.\\ 

The spectral and flux variability, close to the time scale of a day for black-hole mass of $10^8 M_{\odot}$ but covering a wide frequency range, makes things even more confused if we are not able to separate the contribution of the different spectral components. 
For example,  the
X-ray spectral shape of the continuum  is expected to harden (the photon index $\Gamma$
decreases) when the corona temperature (hence the high-energy cut-off)
increases (e.g.  \citealt{haa97}). Such variations in the X-ray spectral shape may be produced by intrinsic
changes in the corona properties (for example, changes in the heating
process efficiency) and/or by variation in the environment like
changes of the soft-photon flux (and consequently of the corona cooling)
produced by, e.g., the accretion disk. 

The origin of the soft X-ray excess has been strongly debated. It was rapidly proposed that it could result from warm Comptonization in an optically thick plasma (e.g. \citealt{wal93,mag98}, and see more recently \citealt{don12}). In this case, a correlation between the UV and soft X-ray emission is expected. If it is produced by blurred ionized reflection, as proposed by, e.g., \cite{cru06}, then the soft X-rays should be linked with the hard ($>$ 10 keV) X-rays and reflection components; i.e., a broad iron line and a bump at $\sim$ 30 keV, are expected. Other authors have explained the soft X-ray excess by smeared absorption (e.g. \citealt{gie06}), but the observations apparently require blurring effects that are too large to be realistic \citep{sch09}. Finally, the effect of partial covering ionized absorption can also account for part of the observed soft excess without requiring material to be moving at extreme velocities (e.g. \citealt{mil08} and the review of \citealt{tur09}).

In summary, it is obvious that improving our understanding of Seyfert galaxy high-energy emission requires long and  broad-band monitoring.
\\

Mrk 509 is a Seyfert 1 galaxy that harbors a super massive black hole of $1.4\times 10^8$ solar masses \citep{pete04}. It was the object of an intense optical/UV/X-rays/gamma-rays campaign in late 2009, with the participation of five satellites (\xmm, INTEGRAL, Swift, Chandra, and HST) and ground-based telescopes (WHT and PAIRITEL). The source characteristics and the whole campaign are explained in \cite{kaa11a}. The main objective of this campaign was to study in detail the physical properties and structure of the warm-absorber outflows in Mrk 509. The heart of the monitoring are ten simultaneous \xmm/INTEGRAL observations, with one observation every four days (60 ks with \xmm\ and 120 ks with INTEGRAL). This paper is part of a series focusing on different results obtained with the campaign. For example, the study of the stacked 600 ks RGS spectrum enabled us to put strong constraints on the outflow nature and geometry \citep{det11,kri11,kaa12}.\\ 

Closely related to the present paper, Mehdipour et al. (\citeyear{mehd11}, hereafter M11) have studied the evolution of the UV/X-ray broad-band emission of Mrk 509 using the OM and pn instruments. M11 show that the UV and soft X-ray continuum emission are tightly correlated, suggesting a common origin for the two continuum emission components. This favors an origin through Comptonization instead of blurred absorption or emission, as discussed above. This interpretation has recently been suggested for several narrow-line Seyfert galaxies \citep{dew07,mid09,jin09}.
The UV/soft X-ray correlation observed in Mrk 509 agrees with the soft X-ray emission being the upper-energy part of the Comptonization of the UV photons in a warm ($kT\sim$ 1 keV) and optically thick ($\tau\sim$ 10-20) corona. It is worth noting that these results do not rule out the presence of other components in the soft X-rays (such as blurred reflection or absorptions). But they do not dominate the variability on a time scale of the few days observed during this monitoring.\\

The study done by M11 was limited to data below 10 keV, and the primary continuum was supposed to be a simple power law. It is known, however, that real Comptonization spectral shapes may be significantly different from a power law (e.g. \citealt{haa93,sve96,pet00}).
Moreover, the interpretation of the spectral variability in terms of
physical quantities, such as the temperature and optical depth of the
corona, is not straightforward from the photon index and high energy cut-off of a phenomenological cut-off power-law
approximation, so it requires the use of approximate formulae to estimate these parameters. \\

In the present paper we aim at testing realistic Comptonization models in a similar way to the Petrucci et al. (\citeyear{pet04}) analysis of the simultaneous IUE/RXTE campaign of NGC 7469. The main advantages of the Mrk 509 campaign are that the data cover the X-ray range from the soft X-ray bands up to 200 keV, while the RXTE energy range, in the case of NGC 7469, was limited to 3-20 keV. Furthermore, the \xmm\ data allow us to put strong constraints on the optical/UV continuum and the iron line complex and to analyze the WA structure precisely. Thus, for the first time, we are able to do a  broad-band spectral analysis of UV to X-ray data of a Seyfert galaxy, on a month's time scale, including all the major spectral components known to be present in this class of objects.\\

The paper is organized as follows. We briefly describe the observation and data reduction in Sect. \ref{obs}, referring the reader to the other papers of the collaboration when appropriate. In Sect. \ref{lc}, we present light curves in different energy bands and interband correlations including the INTEGRAL data. In Sect. \ref{pca}, a principal component analysis of the UV/X-ray/$\gamma$-ray data is shown. The detailed spectral analysis is done in Sect. \ref{specana}. The results are discussed in Sect. \ref{discu} and we state the conclusions in Sect. \ref{conc}.

\begin{figure}
\includegraphics[width=\columnwidth,height=12cm]{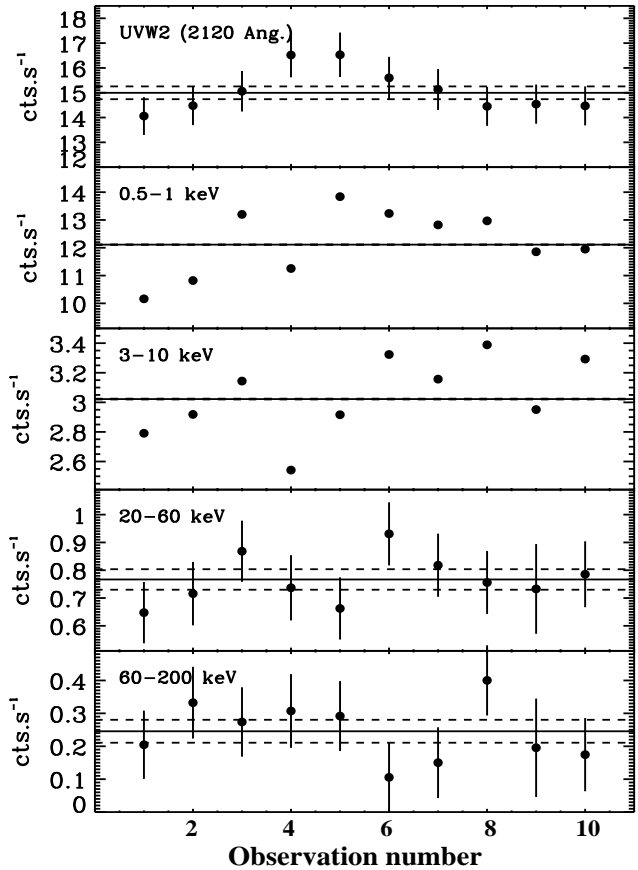}
 \caption{Count rates light curves in different energy ranges. From top to bottom:  2120 \AA, 0.5-1 keV, 3-10 keV, 20-60 keV, and 60-200 keV. The horizontal solid lines are the average values and the horizontal dashed lines represent the corresponding $\pm$ 1$\sigma$ uncertainties.}
\label{figlc}
\end{figure}

\section{Observations and data reduction}
\label{obs}
For details of the observations of \xmm, INTEGRAL, and Swift, precisely please see \cite{kaa11a}. The \xmm/INTEGRAL monitoring started on October 18, 2009 and ended on November 20, 2009. During this campaign, we also obtained UVOT, as well as Swift X-ray, Chandra, and HST/COS data. We further more use archival FUSE data. The first observation, with Swift, started on September 4, 2009, for a total of 19 pointings, and the last observation, with Chandra, ended on December 13, 2009. The HST/COS data were taken about 20 days after the end of  the simultaneous \xmm/INTEGRAL observations and consist of ten HST orbits (see \citealt{kaa11a} for a detailed description of the campaign). The archival FUSE data were obtained October 15, 2000, i.e. about nine years before our campaign.\\

We refer the reader to \cite{pon12b} (P12 hereafter) for details about the \xmm/pn data reduction and especially the gain calibration and correction. This data reduction was done with the latest version of the SAS software (version 10.0.0), starting from the ODF files. The contribution from soft proton flares was negligible during the 2009 monitoring.
The \xmm/OM data treatment and the host galaxy subtraction are detailed in M11, and we also refer the reader to this paper for more information. INTEGRAL/ISGRI data were reduced with OSA 8, with additional ghost-cleaning improvements provided in OSA 9. The presence of SPI annealing meant that 40\% of the observations cannot be used for the SPI instrument. This unfortunately results in too low a signal-to-noise ratio for the SPI data to be used. Observation 6 was split in two owing to operational contingency, however we added both parts to improve statistics.\\

The total \xmm/pn spectrum and the production of the corresponding response and ancillary files were performed with the {\sc MATHPHA}, {\sc ADDRMF}, and {\sc ADDARF} tools within the {\sc HEASOFT} package (version 6.8). The total INTEGRAL spectrum, as well as the corresponding response and ancillary files, were generated using the {\it OSA spe\_pick} tool. The total on-source exposure time is $\sim$ 430 ks for \xmm/pn and $\sim$ 620 ks for INTEGRAL. \\


\section{Light curves and flux correlation}
\label{lc}
The \xmm\ and INTEGRAL count rate light curves are reported in Fig. \ref{figlc} for different energy ranges: at 2120 \AA\, i.e., the wavelength of the OM UVW2 filter, 0.5-1 keV and 3-10 keV for \xmm/pn, and 20-60 keV and 60-200 keV using INTEGRAL ISGRI data. A peak-to-peak variability of 30-40\% is visible in the \xmm/pn light curves and is also detected in the UV band (see also M11).  Some variability seems to be present in the 20-60 keV and 60-200 keV band of ISGRI. Noticeably, the INTEGRAL 20-60 keV light curve looks similar to the \xmm\ 3-10 keV one. However, given the error bars, the ISGRI soft- and hard-band count rate light curves are formally consistent with being constant.\\
\begin{figure}
\includegraphics[width=\columnwidth,height=8cm,angle=0]{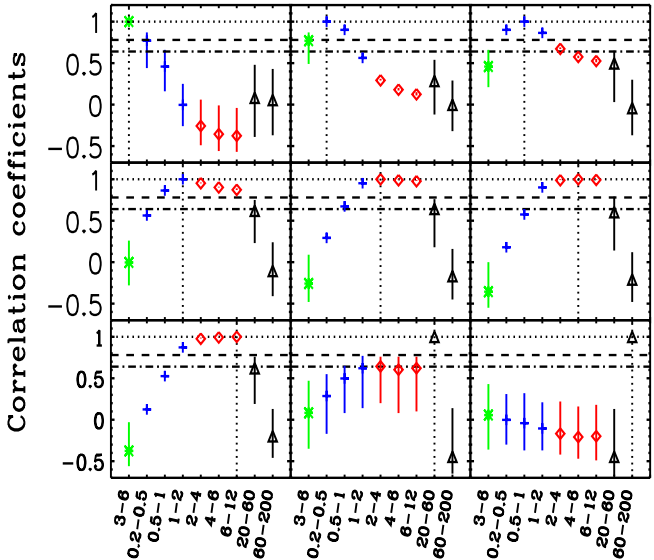}
 \caption{Count rate-count rate correlation coefficients between different energy bands for the 10 observations. In each panel, the green star is the \xmm/OM light curve in the 3-6 eV band; the three blue crosses correspond to the \xmm/pn soft X-ray bands: 0.2-0.5 keV, 0.5-1 keV, 1-2 keV; the three red diamonds correspond to the \xmm/pn medium X-ray bands: 2-4 keV, 4-6 keV, 6-12 keV; and the two black triangles correspond to the INTEGRAL/ISGRI hard X-ray bands: 20-60 keV and 60-200 keV. In each panel, the light curves are correlated with the one marked with the vertical dotted lines. We have overplotted the 95\% and 99\% confidence levels for the significance of the correlation (it corresponds to correlation coefficients of 0.64 and 0.78, respectively, for 10 points) with horizontal dot-dashed and dashed lines, respectively.}
\label{corr}
\end{figure}

An important result observed during this campaign is the strong correlation between the UV and soft X-ray ($<$0.5 keV) flux, while no correlation was found between the UV and the hard ($>$3 keV) X-rays (see M11). This suggests that the UV and soft X-ray excess variability observed on the timescale of the campaign is produced by the same spectral component, possibly thermal Comptonization (see below). \\

We have plotted in Fig. \ref{corr} the linear Pearson correlation coefficients between light curves in different energy ranges, including those above 10 keV from the INTEGRAL/ISGRI instrument. The error bars reported in this figure were estimated in the following way. To simplify, we focus on the error of the Pearson correlation coefficient of 0.78 obtained between the 3-6 eV and 0.2-0.5 keV count rate light curves. For both energy ranges, we simulated 200 light curves of ten count rates, each of these count rates being drawn from a Gaussian distribution probability centered on the {\it observed} values (in these energy ranges) and with a standard deviation equal to the  68\% count rate error of the {\it observed} data. 

We take the example of the simulation of the faked 3-6 eV count rates for Obs 1. The observed count rate for Obs 1 is 14.06 cts s$^{-1}$ with a 90\% error of $\pm$0.76 cts s$^{-1}$. Thus we draw 200 faked 3-6 eV count rates  from a Gaussian centered on 14.06  cts s$^{-1}$ and with a standard deviation of 0.76/1.65=0.46 cts s$^{-1}$. We proceed in the same manner for the 0.2-0.5 keV energy band. Then we correlate the 200 simulated 3-6 eV light curves with the 200 simulated 0.2-0.5 keV light curves to obtain a set of 200$\times$200 fake correlation coefficients. We define  the correlation coefficient $c_{min}$ (respectively $c_{max}$) as the value below (respectively above) which we have 5\% of the simulated correlation coefficients. In the present example, $c_{min}=0.47$ and $c_{max}=0.85$, and we use $c_{min}$ and $c_{max}$ as the corresponding 90\% error on the {\it observed} correlation coefficient. We apply the same procedure for the other correlation coefficients shown in Fig.  \ref{corr}. We have also overplotted, in this figure, the 95\% and 99\% confidence levels for the significance of the correlation. This correspond to correlation coefficients of 0.64 and 0.78, respectively, for 10 points.\\ 

Figure \ref{corr} shows that the  UV and soft X-rays ($<$1 keV) bands are indeed well correlated even if the errors on the correlation coefficient are relatively large. But, in contrast, the absence of correlation with the higher energy bands points to separated components between the low ($<$1 keV) and high ($>$1 keV) energies. Noticeably, the \xmm\ energy bands above $\sim$2 keV appear to vary in unison, and also correlate well with the 20-60 keV INTEGRAL energy band, suggesting a common physical origin.



\begin{figure}
\includegraphics[width=\columnwidth,angle=0]{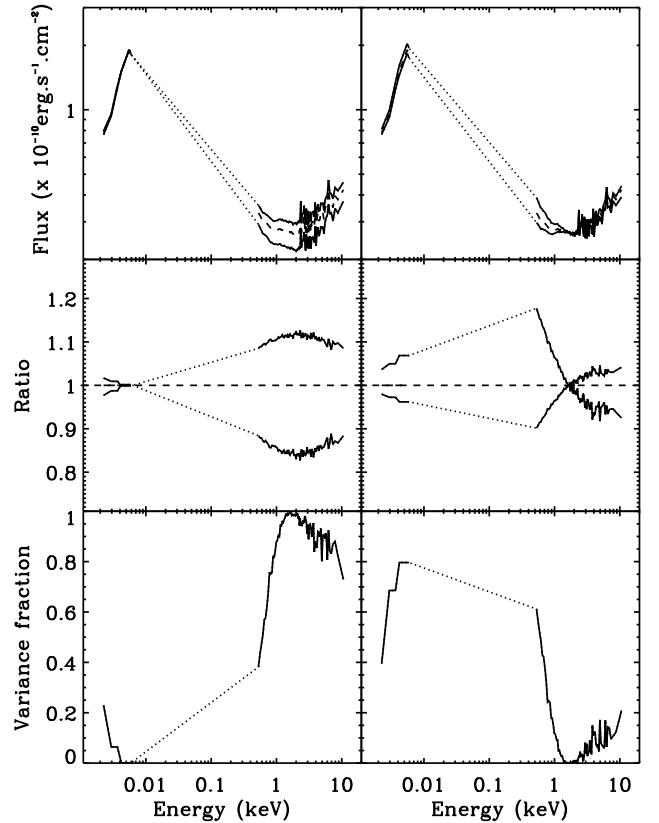} 
 \caption{The first (left panels) and second (right panels) principal components of variability. The upper
  panels show 
  the spectra corresponding to the maximal and minimal coordinate  $\alpha_{j,k}$ over the j=10 observations of the first (k=1,left) and second (k=2, right) eigenvectors.  The middle panels show the ratio of the
  maximum and minimum spectra to the total spectrum of the source. The bottom panels show
  the contribution of each component to the total variance as a function
  of energy. \label{pcafig}}
\end{figure}
\section{Principal component analysis}
\label{pca}
Before going deeper into the spectral analysis, we perform a principal component analysis (PCA) to search for variability
patterns. PCA is a powerful tool for multivariate data analysis which is now widely used in astronomy (e.g. \citealt{pal96,vau04,mal06,mil07,par11}). 
 It allows transforming a number of (potentially) correlated variables into fewer uncorrelated variables called principal components. In the present case, the long exposures per observation enable the use of a large number of energy bins without being limited by the Poisson noise. We choose a number of 91 energy bins, from the UV (6 bins) up to the X-rays (85 bins), with the INTEGRAL data left out due to their low statistics. Consequently, we have {\it p=}10 spectra obtained at different times $t_1$, $t_2$, ..., $t_p$, and binned into {\it n=}91 energy bins $E_{i=1,n}$. Thus we have a $p\times n$ matrix $\mathbf{F}$ whose coefficients are given by the energy fluxes of each spectrum in each energy bin. The PCA gives the eigenvalues and eigenvectors $V_{k=1,n}(E)$ of the correlation matrix of $\mathbf{F}$, thus defining a new coordinate system that best describes the variance in the data. The observed spectra $\mathbf{F}(E,t_j)$ can be decomposed along the eigenvectors with a given set of coordinates $\alpha_{k}$ following $\displaystyle F(E_i,t_j)=\sum_{k=1}^{n}\alpha_{k}(t_j)V_{k}(E_i)$ for each energy bin $E_{i=1,n}$. The coordinates $\alpha_{k}$ vary from observation to observation and their fluctuations account for the observed variance. On the contrary, the eigenvectors are constant and define the different variability modes of the set of observations.\\

The first PCA components are those that reproduce most of the observed variability, with the others dominated by the statistical noise in the data. We have plotted the two first principal components of our analysis in Fig. \ref{pcafig}. The first component (left panels of Fig. \ref{pcafig}) consists mainly in a variability mode dominated by flux variations in the hard energy range ($>$ 1 keV). Most of
the sample variance ($>$87\%) is produced by this component. The second PCA component (right panels of Fig. \ref{pcafig}) is dominated by variability below 1 keV, and represents 10\% of the total variance. The 4\% of variance left is distributed between the higher order components not shown here. This result nicely supports the idea that two main components are dominating the broad-band variability of this campaign, one in the UV-soft X-rays and the other in the hard X-rays.

\section{Spectral analysis}
\label{specana}
Here we study the spectral energy distribution between the optical, UV, X-ray, and gamma-ray bands for Mrk 509. The \xmm/pn spectra have been rebinned in order to have a minimum of 25 counts per bin and three bins per resolution element. The cross calibration normalization between \xmm\ and INTEGRAL is fixed to one. All spectral fits were performed using the \xspec\ software (version 12.3.0).
In Sect. \ref{above3keV}, the reported errors are at the 90\% confidence level for one interesting parameter, while in Sect. \ref{comptmodel}, owing to the potential strong correlations between the model parameters, the reported errors are at the 90\% confidence level for two interesting parameters.

\subsection{Above 3 keV}
\label{above3keV}
We first concentrate our analysis on the data above 3 keV, i.e. on a domain known to be dominated by the primary continuum. This agrees with the PCA analysis (Sect. \ref{pca}), which suggests that a unique component dominates the variability in this energy range. In this section, we are interested in the global spectral behavior of the source. We thus simply use a cut-off power-law for the continuum. More realistic Comptonization models are used in Sect. \ref{comptmodel} when the broad-band UV/X-rays/gamma-rays energy range is used.

To start, we analyze the total spectrum over the entire campaign to determine the best model components to be used. Afterwards we apply the corresponding model to the different observations in Sect. \ref{allpointings1}. 

\subsubsection{The total spectrum}
\label{averagespectrum}
\begin{figure}
\includegraphics[height=\columnwidth,angle=-90]{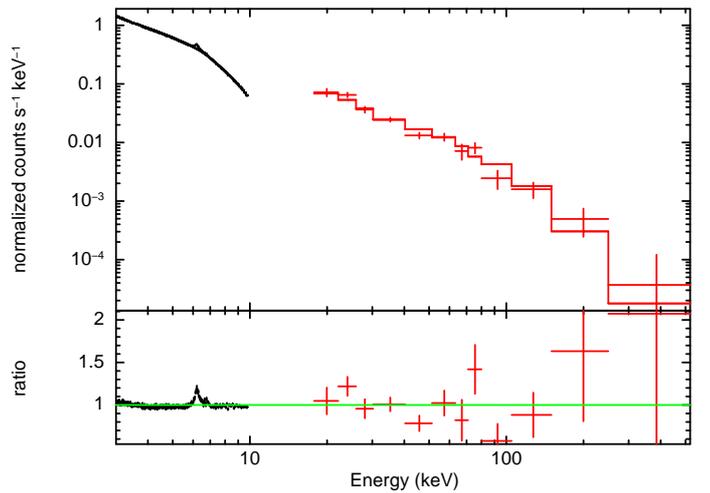}
 \caption{{\bf Top:} Total \xmm\ (pn, black) and INTEGRAL (ISGRI, red) data above 3 keV. The solid line is the best-fit power-law model.  {\bf Bottom:} Ratio data/model. The iron line is clearly visible and no high-energy cut-off is required.}
\label{average_fitpo}
\end{figure}

The best fit and data/model ratio of the total spectrum are plotted in Fig. \ref{average_fitpo}. It clearly shows the presence of an iron line near 6.4 keV. The presence of a high-energy cut-off is not required by the data, and a value  below $\sim$200 keV is even excluded  at the 90\% confidence.\\ 

The iron line complex observed during our campaign is discussed in detail in P12. It can be decomposed into a narrow core ($\sigma\simeq$ 0.03 keV) and a broader component ($\sigma\simeq$ 0.22 keV). When looking to archival (\xmm, Chandra, and Suzaku) data, the narrow core appears to be constant on a several-year timescale, peaking at an energy consistent with 6.4 keV. This suggests an origin in remote neutral material (see also \citealt{der04,pon09}). On the contrary, the intensity of the broader component seems to follow the variations of the continuum throughout the campaign (i.e. on a timescale of a few days), without a measurable time lag 
(see Fig. 4 of P12). Given the sampling of our monitoring, this suggests an origin within a few light days from the black hole, the width of the line being compatible with a Keplerian motion at 40--1000 gravitational radii from the 1.4$\times 10^8\ M_{\sun}$ supermassive black hole of Mrk 509. Note that the relatively large EW of $\sim$50 eV of this component
suggests that part of it could originate in the inner BLR \citep{cos12}.\\ 

\begin{table*}
\begin{center}
\begin{tabular}{ccccccccc}
\hline
Obs. & $\Gamma$ & $E_c$ (keV) & $E_{line}$ (keV) & $\sigma_{line}$ (eV) & $F_{line}$ & Line EW (eV) & $N_{pexmon}\ (\times 10^{-3})$ & $\chi^2 /dof$\\
\hline
Total & 1.74$\pm$0.01& $>$1000 & 6.58$\pm$0.04 & 300$\pm$50 & 2.9$\pm$0.4 & 55$\pm$20 & 2.9$\pm$0.3 & 468/243\\
Obs. 1 & 1.71$\pm$0.04& $>$90 & 6.60$\pm$0.15 & 370$_{-110}^{+150}$ & 3.0$\pm$1.0 & 60$_{-5}^{+30}$ & (f) & 213/251\\
Obs. 2 & 1.76$\pm$0.02& $>$310 & 6.82$_{-0.15}^{+0.10}$ & 510$_{-150}^{+230}$ & 4.4$_{-0.7}^{+1.9}$ & 100$_{-70}^{+80}$ & (f) & 278/251\\
Obs. 3 & 1.76$\pm$0.04& $>$300 & 6.55$\pm$0.10 & 210$_{-80}^{+140}$ & 2.5$\pm$1.0 & 40$\pm$30 & (f) & 287/251\\
Obs. 4 & 1.74$\pm$0.03& $>$250 & 6.70$\pm$0.20 & 660$_{-190}^{+350}$ & 4.8$_{-1.8}^{+2.8}$ & 110$_{-70}^{+90}$  & (f) & 270/251\\
Obs. 5 & 1.75$\pm$0.03& $>$100 & 6.57$\pm$0.15 & 360$_{-130}^{+200}$ & 3.2$_{-1.2}^{+1.5}$ & 60$_{-40}^{+30}$  & (f) & 264/251\\
Obs. 6 & 1.73$\pm$0.03& $>$120 & 6.52$\pm$0.05 & $<$140 & 1.8$_{-0.6}^{+0.8}$ & 30$_{-15}^{+20}$  & (f) & 300/251\\
Obs. 7 & 1.77$\pm$0.02& $>$350 & 6.50$\pm$0.10 & 300$_{-120}^{+140}$ & 3.0$_{-1.0}^{+1.1}$ & 50$_{-40}^{+60}$  & (f) & 310/251\\
Obs. 8 & 1.76$\pm$0.02& $>$200 & 6.60$\pm$0.10 & 250$_{-90}^{+120}$ & 3.4$_{-1.0}^{+1.6}$ & 60$_{-40}^{+30}$  & (f) & 256/251\\
Obs. 9 & 1.76$_{-0.04}^{+0.02}$& $>$160 & 6.51$\pm$0.09 & 330$_{-90}^{+100}$ & 4.0$_{-1.0}^{+1.2}$ & 80$_{-30}^{+40}$  & (f) & 281/251\\
Obs. 10 & 1.68$\pm$0.04 & $>$80 & 6.54$\pm$0.08 & 170$_{-60}^{+90}$ & 2.3$_{-0.7}^{+1.0}$ & 80$_{-30}^{+20}$  & (f) & 226/251\\
\hline
\end{tabular}
\end{center}
\caption{Best-fit parameters when fitting the data above 3 keV. A cut-off power law is used for the primary continuum, a Gaussian for the broad-line component, and a neutral reflection (\pexmon) for the narrow iron line and the reflection bump. For the individual observations, the \pexmon\ parameters are fixed (f) to the best-fit values of the total spectrum. \label{tabge3keV}}
\end{table*}

Following P12, we fit the iron line profile with two model components. For the narrow core we used the \pexmon\ model of \xspec\ \citep{nan07}. It reproduces the reflection component following the {\sc pexrav} model \citep{zdz91} of \xspec, but in addition it includes the K$\alpha$ and K$\beta$ neutral iron line, the Compton shoulder, and the nickel K$\alpha$ line, all these components being computed ``self-consistently'' according to the \cite{geo91} simulations. We fixed the inclination angle to 30 deg. in \pexmon. Since it is supposed to model remote reflection, the illuminating spectrum should be an average of the different states of the central source. Thus, we chose to fix the illuminating power-law photon index to 1.9 and a fixed high-energy cut-off at 300 keV. The exact values of these parameters have no significant effects on the results shown in this paper. We let only the normalization of \pexmon\ free to vary in the fit of the total spectrum. 
Concerning the broader component we used a simple Gaussian whose energy, flux and width are left free to vary during the fit.\\ 

The best-fit parameter values for the total spectrum are reported in the first row of Table \ref{tabge3keV}. The power-law photon index $\Gamma=1.74\pm 0.01$,  the broad line component peaks at  6.58$\pm$0.04, and has an EW of 55$\pm$20 eV and the \pexmon\ normalization is $N_{pexmon}=(2.9\pm 0.3) \times 10^{-3}$. The \pexmon\ normalization is mainly constrained by the intensity of the core of the narrow iron-line component. The fit is not satisfactory ($\chi^2$/dof=468/243), mainly due to the residuals at low energies (close to 3 keV) and possibly due to the influence of the soft X-ray excess. 

\subsubsection{Individual observations}
\label{allpointings1}
We applied the same fitting procedure to the ten different observations. The only varying parameters are the photon index and normalization of the power law and the high-energy cut-off and the characteristics (energy, width, and intensity) of the broad-line component. The \pexmon\ normalization is not expected to vary during the campaign (as the signature of remote reflection) and is fixed to the best-fit value obtained with the total spectrum. 

The best-fit parameters for each pointing are reported in Table \ref{tabge3keV}. For the individual pointings, we always obtained an acceptable fit. The best-fit photon index is only marginally variable with a range in between 1.68 and 1.77, and the presence of a high-energy cut-off is never required. The shape of the high-energy spectra of Mrk 509 thus changes only slightly during the monitoring. The broad-line component is always statistically needed. The line equivalent width is consistent with a constant with an average value of 55$\pm$ 10 eV. The line flux follows the continuum flux variability, in agreement with the results obtained by P12. However, its energy is only marginally consistent with the 6.4 keV neutral energy, possibly indicating the presence of ionized iron (see P12 for a more detailed discussion).

\subsection{From UV to hard X-rays/soft gamma-rays Comptonization models}
\label{comptmodel}
By fitting above 3 keV we ignore the spectral complexity of the soft X-ray range where soft X-ray excess and warm-absorber (WA) features are present. These spectral features have to be included in the fitting procedure of the whole dataset from UV to hard X-rays/soft gamma-rays. They are discussed in some detail in the next paragraphs. \\

It is worth noting, at this stage, that the purpose of the present paper is not to test different  models (especially concerning the soft X-ray excess or the primary continuum).  We instead make the choice of fiting a few, physically motivated spectral components, which will permit us to discuss the corresponding physical interpretation of the data in depth.
 
\subsubsection{The model components}

\paragraph{The warm absorber:} the WA has been comprehensively analyzed by \cite{det11}, and we refer the reader to this article for more details. We produced a table from the best model of the WA derived from this analysis and included this table in \xspec.

\paragraph{The soft X-ray excess:} as already explained in the introduction, different models exist in the literature to interpret this component. The most commonly used in the last few years are blurred ionized absorption or reflection and warm Comptonization. That no evidence for changes in the X-ray WA was found in Mrk 509 despite a soft X-ray intensity increase of $\sim$60 \% in the middle of our campaign \citep{kaa11a} disfavors ionized absorption as the origin of the optical-UV/soft X-ray correlation. 

Concerning ionized reflection, it is known to be able to produce a soft X-ray excess \citep{cru06,pon09,cer11,nod11}. Moreover, since part of the X-ray emission is reprocessed in opt-UV in the disk, we also expect some correlation between these energy bands. Two arguments weaken, however, the importance of ionized reflection in the opt-UV/X-ray variability behavior observed in Mrk 509. First we expect a reflection bump peaking near 10 keV that should vary with the soft part of the reflection. But no correlation is observed between the soft and the hard X-ray bands (see Fig. \ref{corr}). It is true, however, that this could be partly explained by the low statistics above 10 keV. But more importantly, part of the iron line should also be produced by this reflection component and would be expected to vary with the soft X-rays. However, neither the narrow core of the line nor its broader component show variations that agree with those observed in the soft X-rays (see Sect. \ref{averagespectrum}).\\

Following M11, we chose to model the UV-soft X-ray emission by thermal Comptonization. We used the \nthcomp\ model of \xspec. \nthcomp\ covers a wide range of parameters and is well adapted to warm (kT$\sim$ 1 keV)\footnote{In \xspec, the lower limit of the temperature parameter of \nthcomp\ is 1 keV. However, this can be modified and put to a lower value. We have checked that the model still works for kT$\sim$0.5 keV by comparing its spectral shape with the one obtained with \eqpair.}  and optically thick ($\tau\sim$ 10--20) plasma (i.e. what we call a {\sc warm} corona in the following), such as the one expected to reproduce the steep soft X-ray emission. The free parameters of \nthcomp\ are the temperature of the corona \twc\ \footnote{The letters ``wc'' means warm corona and refer to all the parameters of the \nthcomp\ model.}, the soft-photon temperature \tbbwc , and the asymptotic power-law index \gwc\ of the spectrum. We prefer to use this model instead of {\comptt} (which is similar to the {\sc comt} model in {\sc spex} used by M11), because  the seed photons in  {\sc comptt} only have a black-body distribution. With \nthcomp\ we can choose a multicolor-disk distribution. It is also fast to use in \xspec\ (see also App. \ref{optdepth} for a comparison of these two models). 

Note however that \nthcomp\ does not provide the corona optical depth directly. It has to be deduced in another manner, generally by comparing its spectral shape to another model that provides the optical depth.   This is, however, a possible source of uncertainties and is discussed in detail in App. \ref{optdepth}. In the following, the optical depth of the {\sc warm} corona is deduced from a comparison with the \eqpair\ model.

%
%
%


\paragraph{The primary continuum:} concerning the high-energy continuum, the cut-off power-law model is known to be a poor approximation of realistic thermal Comptonization models. First, the high-energy cut-off shape of Comptonization spectra differs from an exponential cut-off. Besides, Comptonization spectra result from the sum of multiple bumps, each one corresponding to the different Compton scattering orders experienced by the soft seed photons in  the corona. For small optical depth, the spectral shape is bumpy. Finally, Comptonization spectra naturally include the seed photons  that cross the corona but are not Compton-scattered (the so-called zeroth scattering order).\\

We used the Comptonization model \compps\ \citep{pou96} of \xspec\ to model the high-energy primary continuum. \compps\ is well adapted to reproducing the emission of a hot and optically thin plasma, i.e., what we call the {\sc hot} corona in the following. The fit parameters of \compps\ are the temperature 
of the corona $kT_{hc}$\footnote{The letters ``hc'' mean hot corona, and refer to all the parameters of the \compps\ model.} , either the optical depth $\tau_{hc}$ or the parameter $\displaystyle y_{hc}=4\frac{kT_{hc}}{m_ec^2}\tau_{hc}$, where $y_{hc}$ is equivalent to the Compton parameter for small ($<$ 1) optical depths, the temperature of the soft photons produced by the cold phase $kT_{bb,hc}$ (also assuming a multicolor-disk distribution) and the geometry of the disk-corona configuration.

We prefer to use the Compton parameter instead of the optical depth in the fitting procedure  to minimize the model degeneracy. Indeed, the temperature and optical depth are generally strongly correlated in the fitting procedure, with the same 2-10 keV slope obtained with different combinations of these two parameters (see e.g. \citealt{pet01sey}).  We also assume a slab geometry for the corona-disk system.

\paragraph{The iron line:} following the results obtained in Sect. \ref{above3keV}, we use \pexmon\ to fit the narrow core of the iron line and a Gaussian to fit a broad-line component if needed.\\

It is important to point out that no link between the Comptonized spectrum and the soft UV emission is imposed a priori.  The model simply adjusts its parameters, independently of each other, to fit the data. It is only a posteriori that the resulting best-fit values can be interpreted in a physically motivated scenario. \\



\subsubsection{The COS-FUSE data set}
The use of non-simultaneous optical-UV to hard X-ray data could produce spurious constraints on Comptonization models and should be avoided; however, a few arguments make the use of the COS and FUSE data in our spectral analysis reasonable. First, 
the UV continuum shape of Mrk 509 did not change significantly, even for a 60\% lower flux as observed with FOS \citep{kri11}. 
This suggests that the UV spectral shape of Mrk 509 only weakly changes as its flux varies. Moreover, some of the Swift data were obtained simultaneously with the COS observations in a bandpass comparable to the OM bands. The Swift fluxes are at values comparable to OM points near the beginning and the end of the \xmm\ campaign, and only ~12\% below the mean flux of the whole campaign \citep{kri11}. Given the lack of evidence for variation in the spectral shape and the fact that only a small-scale factor is needed to bring the COS data in line with the mean OM fluxes, we decided to use of the COS data in our fit procedure. 

The scaling factor to match the FUSE data with the OM one, still assuming a constant spectral shape, is much larger ($\sim$ 2, see M11 for more information on the method used to rescale the COS and FUSE data) and the use of the FUSE data is more arguable. However, it allows for a better constraining of the maximum of the UV bump. For these different reasons, we chose to include the COS and FUSE data in our spectral analysis.\\

%
%
%


\begin{figure}
\includegraphics[width=\columnwidth,angle=0]{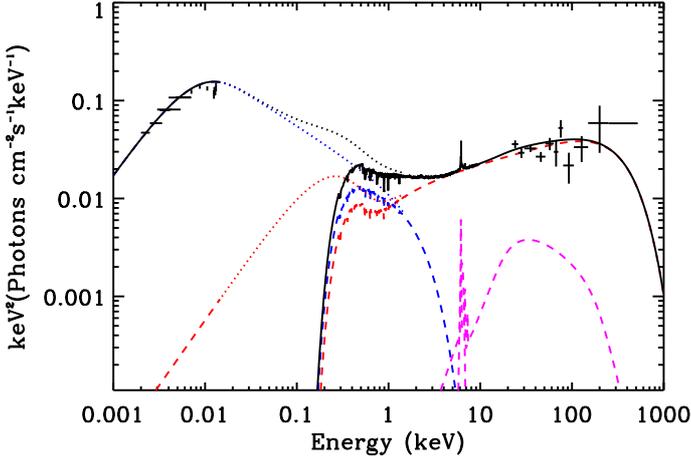}
 \caption{Unfolded best-fit model of the total data spectrum using a thermal Comptonization component (\compps\ in \xspec) for the primary continuum. The data are the black crosses. The solid black line is the best-fit model including all absorption (WA and Galactic). The dashed lines correspond to the different spectral components, \hot\ (in red) and \warm\ (in blue) corona emission as well as the reflection (magenta) produced by \pexmon\ including the effects of absorptions, while the dotted line are absorption free.}
\label{ploteeufcomp}
\end{figure}

\subsubsection{The total spectrum}

\label{avercomp}

\begin{table*}
\begin{center}
\begin{tabular}{cccccccccccc}
\hline
Obs &\gwc\ & \twc\ & \tawc\ & \thc\ & \yhc\  & \tahc\ & \tbbwc\ &\tbbhc\ & $N_{pexmon}$ & $\chi^2 /dof$\\
 & & (keV) &  &(keV)&   &  & (eV) & (eV) &($\times$ 10$^{-3}$)  & \\
\hline
\hline
Total & 2.595$_{-0.005}^{+0.005}$ & 0.56$_{-0.02}^{+0.02}$ & 18.4$_{-0.4}^{+0.3}$& 123$_{-6}^{+5}$ & 0.63$_{-0.01}^{+0.01}$ &  0.65$_{-0.01}^{+0.01}$& 3.4$_{-0.1}^{+0.1}$ &  100$_{-1}^{+1}$& 2.9  & 1387/690\\
Obs 1   & 2.56$_{-0.01}^{+0.03}$ & 0.48$_{-0.04}^{+0.07}$ & 20.3$_{-0.1}^{+1.1}$&   102$_{-10}^{+31}$ & 0.63$_{-0.01}^{+0.06}$ &  0.89$_{-0.03}^{+0.10}$& 2.6$_{-0.1}^{+0.2}$ &    97$_{-7}^{+13}$& (f)  & 396/415\\
Obs 2   & 2.55$_{-0.01}^{+0.01}$ & 0.54$_{-0.06}^{+0.06}$ & 19.3$_{-1.5}^{+0.7}$& 122$_{-17}^{+18}$ & 0.63$_{-0.02}^{+0.02}$ &  0.66$_{-0.08}^{+0.03}$& 2.7$_{-0.1}^{+0.2}$ &    95$_{-2}^{+1}$& (f)  & 510/415\\
Obs 3   & 2.53$_{-0.01}^{+0.01}$ & 0.54$_{-0.06}^{+0.04}$ & 19.3$_{-0.4}^{+1.2}$& 120$_{-19}^{+8}$ & 0.60$_{-0.02}^{+0.02}$ &  0.65$_{-0.01}^{+0.07}$& 2.9$_{-0.1}^{+0.2}$ &  108$_{-2}^{+1}$& (f)  & 501/415\\
Obs 4   & 2.60$_{-0.01}^{+0.01}$ & 0.56$_{-0.04}^{+0.05}$ & 17.1$_{-0.1}^{+2.1}$& 164$_{-15}^{+18}$ & 0.54$_{-0.01}^{+0.02}$ &  0.42$_{-0.03}^{+0.02}$& 3.2$_{-0.5}^{+0.1}$ &    102$_{-2}^{+1}$& (f)  & 500/415\\
Obs 5   & 2.57$_{-0.01}^{+0.01}$ & 0.50$_{-0.03}^{+0.03}$ & 19.8$_{-1.5}^{+1.4}$& 141$_{-13}^{+11}$ & 0.51$_{-0.01}^{+0.01}$ &  0.46$_{-0.03}^{+0.02}$& 3.3$_{-0.3}^{+0.2}$ &  105$_{-1}^{+1}$& (f)  & 467/415\\
Obs 6   & 2.54$_{-0.01}^{+0.03}$ & 0.59$_{-0.08}^{+0.04}$ & 18.5$_{-2.5}^{+1.4}$& 108$_{-21}^{+8}$ & 0.66$_{-0.04}^{+0.03}$ &  0.78$_{-0.03}^{+0.08}$& 2.9$_{-0.2}^{+0.1}$ &  116$_{-2}^{+2}$& (f)  & 518/415\\
Obs 7   & 2.54$_{-0.01}^{+0.01}$ & 0.54$_{-0.08}^{+0.05}$ & 19.0$_{-0.8}^{+1.3}$& 119$_{-20}^{+13}$ & 0.60$_{-0.03}^{+0.02}$ &  0.67$_{-0.02}^{+0.07}$& 2.9$_{-0.2}^{+0.4}$ &  104$_{-2}^{+1}$& (f)  & 513/415\\
Obs 8   & 2.52$_{-0.01}^{+0.02}$ & 0.52$_{-0.03}^{+0.07}$ & 20.0$_{-0.9}^{+3.2}$&   78$_{-10}^{+13}$ & 0.64$_{-0.02}^{+0.05}$ &  1.05$_{-0.12}^{+0.09}$& 2.6$_{-0.1}^{+0.2}$ &  114$_{-3}^{+2}$& (f)  & 494/415\\
Obs 9   & 2.55$_{-0.01}^{+0.02}$ & 0.49$_{-0.04}^{+0.04}$ & 20.2$_{-1.4}^{+0.1}$& 115$_{-16}^{+9}$ & 0.60$_{-0.02}^{+0.02}$ &  0.67$_{-0.02}^{+0.08}$& 3.0$_{-0.2}^{+0.1}$ &  100$_{-1}^{+2}$& (f)  & 450/415\\
Obs 10 & 2.53$_{-0.01}^{+0.01}$ & 0.53$_{-0.05}^{+0.05}$ & 19.7$_{-0.9}^{+1.2}$& 102$_{-12}^{+12}$ & 0.67$_{-0.03}^{+0.03}$ &  0.84$_{-0.05}^{+0.04}$& 2.6$_{-0.2}^{+0.3}$ &  108$_{-4}^{+1}$& (f)  & 409/415\\
\hline
\end{tabular}
\end{center}
\caption{ Best-fit parameters of the total spectrum and the different observations using the Comptonization model \compps\ for the primary continuum and \nthcomp\ for the UV-soft X-rays. The optical depth $\tau_{wc}$ of the {\sc warm} corona is estimated from \eqpair\ fits (see App. \ref{optdepth}).}
\label{tab3}
\end{table*}

First, we assume the same soft-photon temperatures for the two Comptonization models, i.e., \tbbhc=\tbbwc. The fit is very bad with $\chi^2/dof=$8850/691, and the data/model ratio shows several features in the X-rays. Relaxing the constraint of equal soft temperatures between the two plasmas enables us to reach a much better fit with $\chi^2/dof=$1387/690. The corresponding best-fit parameter values are reported in the first row of Table \ref{tab3}. We have plotted in Fig. \ref{ploteeufcomp} the corresponding unfolded best-fit model.\\ 

The two plasmas clearly have different characteristics. The \warm\ corona is optically thick with \tawc$\sim$20, a temperature \twc\ on the order of 600 eV and a soft-photon temperature \tbbwc\ $\sim$3 eV. In contrast, the \hot\ corona is optically thin \tahc$\sim$0.6, with a temperature \thc\ of about 100 keV and a soft-photon temperature \tbbhc$\sim$100 eV. 

In this model, the optical/UV data are entirely fitted by the low-energy part of the {\sc warm} plasma emission (see Fig. \ref{ploteeufcomp}). The soft-photon bump of the \hot\ corona, instead, peaks in the soft X-ray range and contributes to the soft X-ray excess. The soft-photon temperature of $\sim$100 eV suggests that the \hot\ corona is localized in the inner region of the accretion flow compared to the \warm\ corona; however, a 100 eV temperature is large for a ``standard'' accretion disk around a $\sim 10^8\ M_{\sun}$ black hole. This point is discussed in Sect. \ref{discrepro}. \\

The value of 100 keV for the \hot\ corona temperature is compatible with the lower limit on the high-energy cut-off $E_c >$200 keV obtained when fitting with a cut-off power-law shape for the primary continuum (see Sect. \ref{above3keV}). Indeed, as said in the introduction, the cut-off spectral shape of a Comptonization spectrum differs from an exponential cut-off and, roughly speaking, $E_c$ is generally about a factor 2 or 3 larger than the ``real'' corona temperature. Thus $E_c >$ 200 keV agrees with a corona temperature of $\sim$100 keV.\\

The fit of the total spectrum is still statistically not acceptable, however, with the presence of strong features in the data/model ratio, especially in the soft X-rays. The FUSE points are also slightly below the best-fit model (see Fig. \ref{ploteeufcomp}). A possible reason for these discrepancies, given that the fits are reasonably good for each individual observation (see next section), could be the source spectral variability.  The total spectrum is then a mix of different spectral states that cannot be easily fit with ``steady-state'' components. 



\subsubsection{Individual observations}
\begin{figure}
\begin{tabular}{c}
\includegraphics[width=\columnwidth,angle=0]{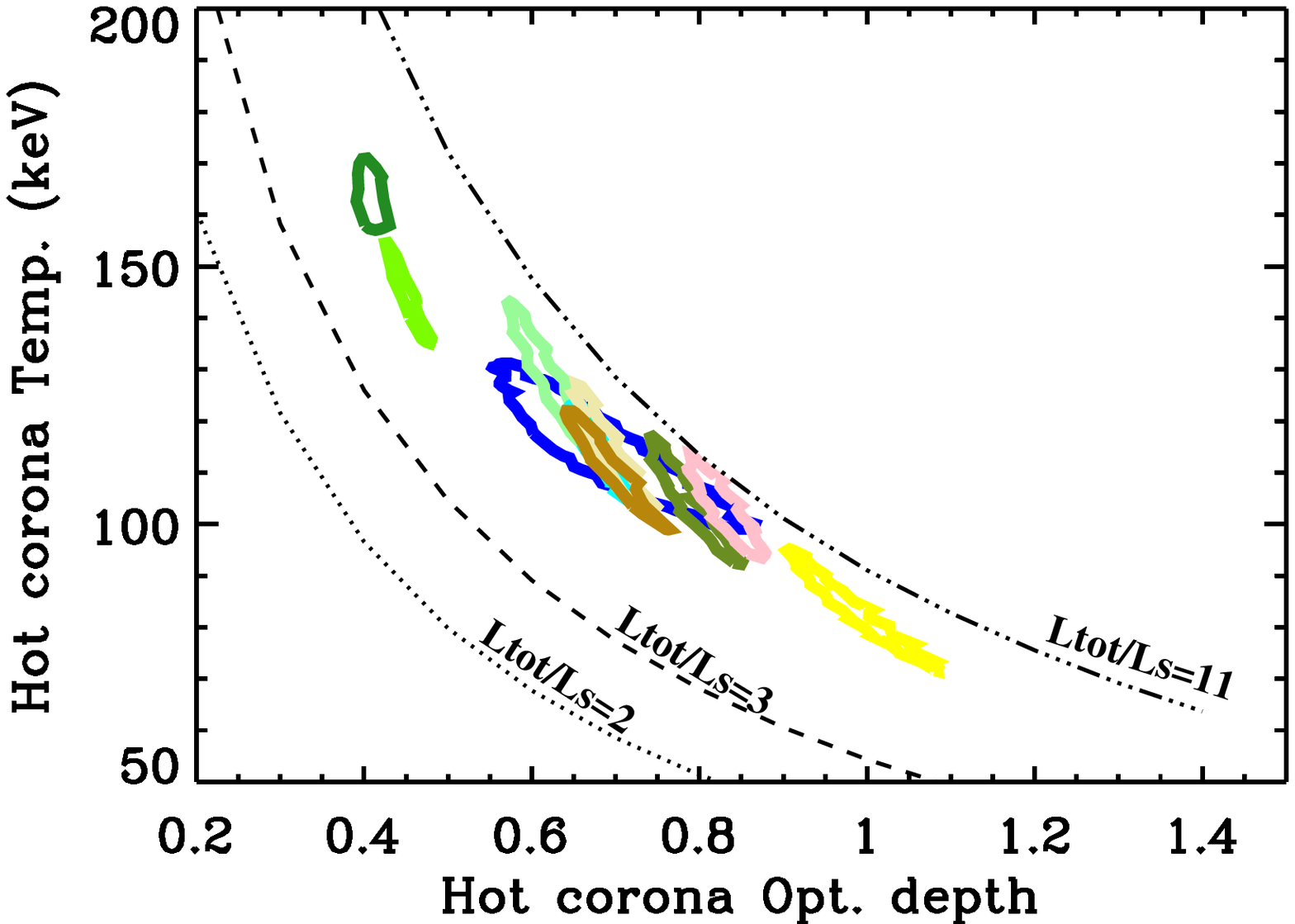}\\
\includegraphics[width=\columnwidth,angle=0]{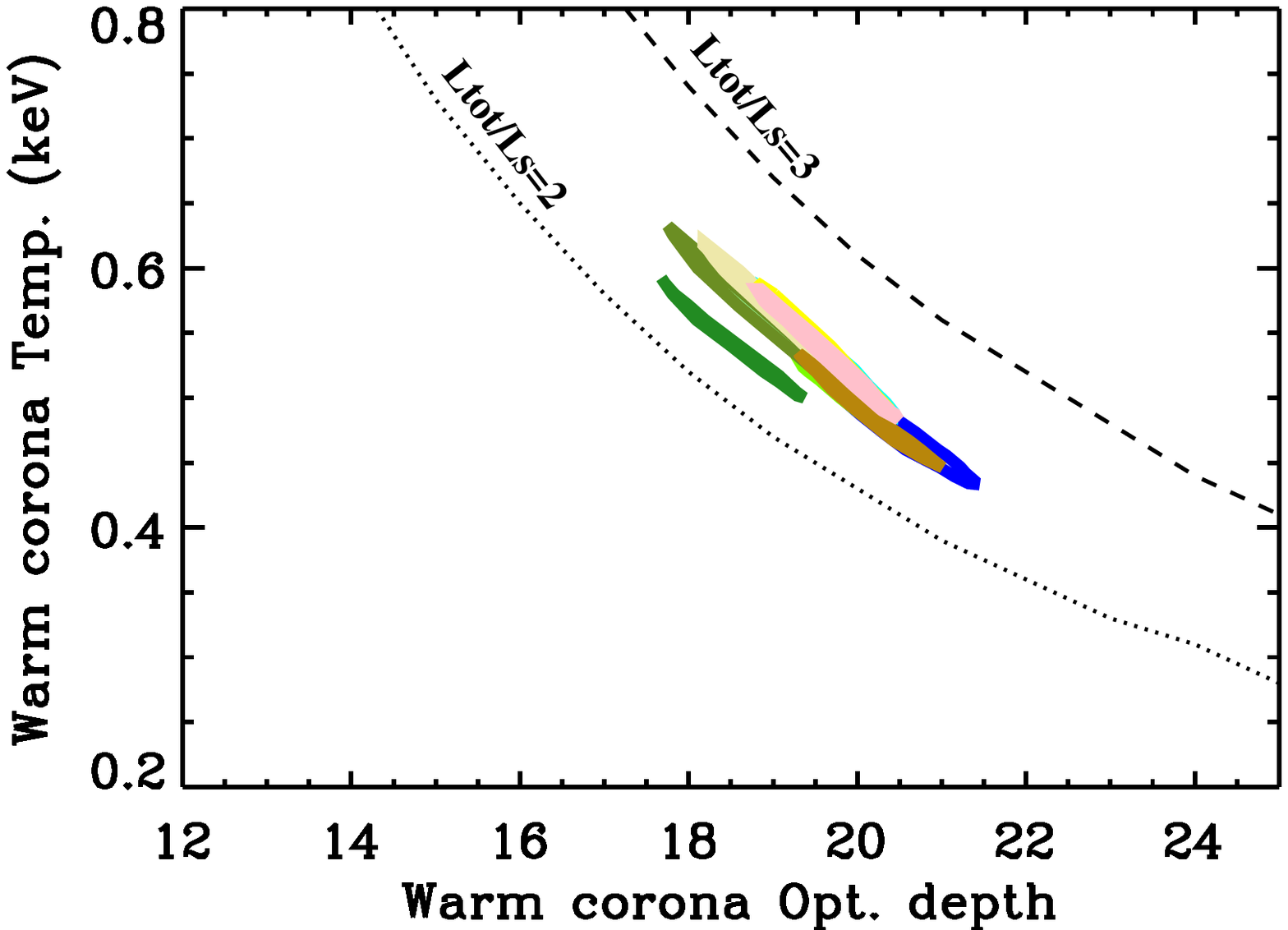}
\end{tabular}
 \caption{Contour plots of the temperature versus optical depth for the {\sc hot} (top) and {\sc warm} (bottom) corona for the 10 observations. We have overplotted the predicted relationship $\tau$ vs. $kT$ in steady state assuming a soft-photon temperature of 10 eV and for different Compton amplification ratios: dotted line: $L_{tot}/L_s$=2; dashed line: $L_{tot}/L_s$=3; dot-dot-dot-dashed: $L_{tot}/L_s$=11. 
 }
\label{plotthetavstau}
\end{figure}

We repeated the same fitting procedure for the ten different observations, letting \tbbhc\ and \tbbwc\ free to vary independently. As in Sect. \ref{allpointings1}, the \pexmon\ normalization is fixed to the best-fit value obtained with the total spectrum. The best-fit parameters for each pointing are reported in Table \ref{tab3}. The 90\% contour plots of the temperature vs. the optical depth are plotted in Fig. \ref{plotthetavstau} for both coronae and for the ten observations.  The errors reported in Table \ref{tab3} are computed from these contour plots and not from the {\it error} command of \xspec. Indeed, the shape of the $\chi^2$ surface is relatively complex for the models used, and confidence intervals are found better by determining the locus of constant $\Delta\chi^2$ than by the use of the error command.\\

\begin{figure}
\includegraphics[width=1.05\columnwidth,angle=0]{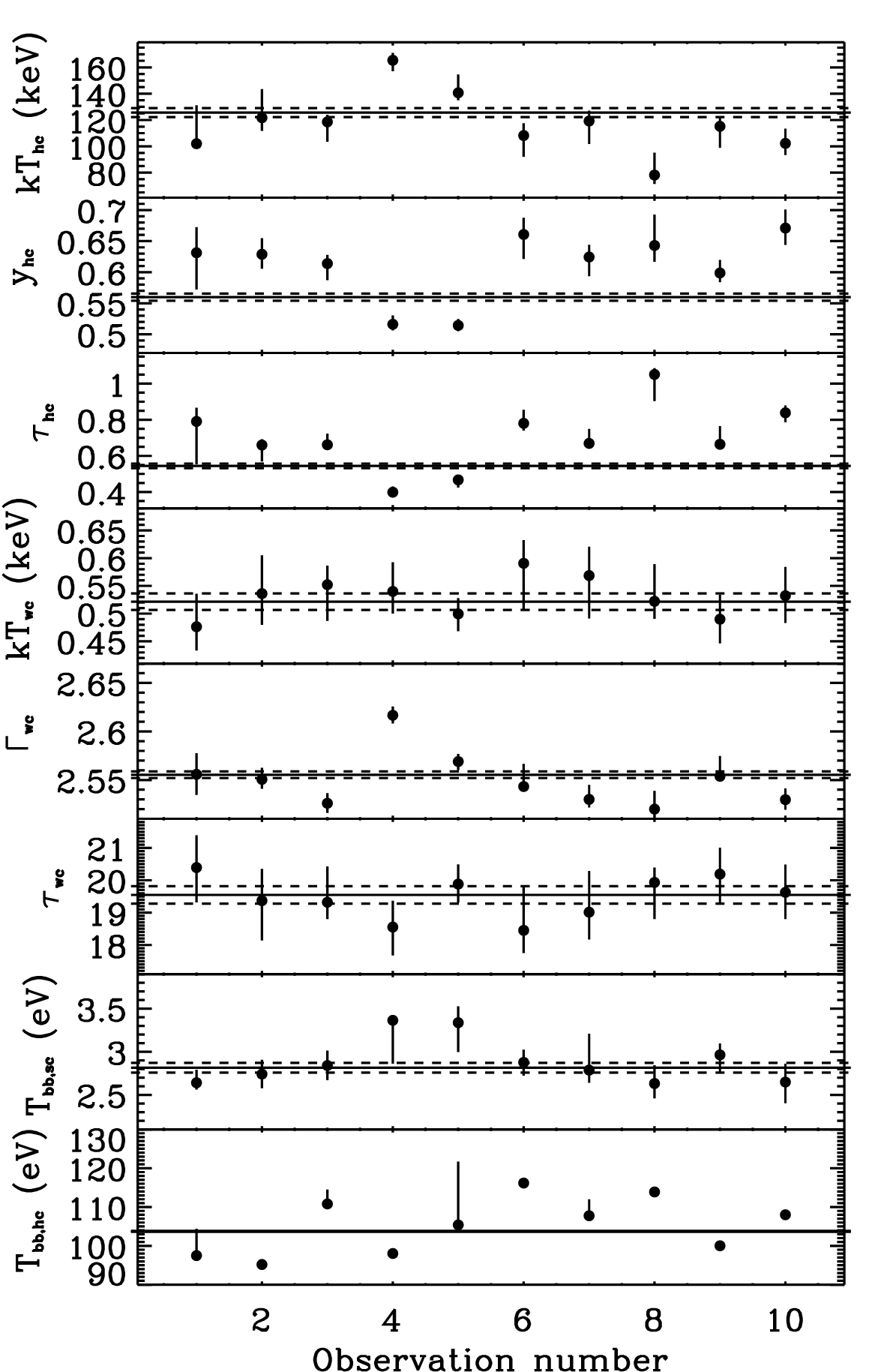}
 \caption{Time evolution of the different fit parameters. From top to bottom: the \hot\ corona temperature \thc, Compton parameter \yhc, optical depth \tahc, and the \warm\ corona temperature \twc, photon index \gwc, optical depth \tawc, and the soft photon temperatures $T_{bb,wc}$ and $T_{bb,hc}$. The solid lines show the best-fit constant values and the solid lines the $\pm1\sigma$ uncertainties.}
\label{lcparam}
\end{figure}

The time evolutions of the different best-fit parameters are plotted in Fig. \ref{lcparam}. For each parameter we have overplotted their weighted average value with the corresponding $\pm1\sigma$ uncertainties. All the light curves are inconsistent with a constant at more than 99\% confidence except for the temperature, optical depth, and soft photon temperature of the \warm\ corona.\\

\subsubsection{Heating and cooling}
We call $L_{tot}$ the total corona luminosity and $L_s$ the {\it intercepted} soft-photon luminosity, i.e., the part of the soft-photon luminosity emitted by the cold phase that effectively enters and cools the corona. The difference $L_{tot}-L_s$ then corresponds to the intrinsic heating power provided by the corona to comptonize the seed's soft photons. The ratio $L_{tot}/L_s$ is called the Compton amplification ratio.\\

From our best fits obtained for each observation, we can estimate the seed's soft-photon luminosity $L_s$ that enters both coronae. Indeed, the number of photons is conserved in the Comptonization process. Consequently, from the number of photons measured in our \nthcomp\ and \compps\ model components and the temperature of the soft-photon field, we can deduce (assuming a multicolor-disk distribution) the corresponding soft-photon luminosity that crosses and cools each corona (see App. \ref{ampl}). In this reasoning, we assume that no extra component participates in the observed opt-UV emission apart the Comptonization spectra. Any extra component would decrease the effective seed soft-photon luminosity $L_s$, implying that our estimates of $L_s$ are strict upper limits. 

The total corona luminosity $L_{tot}$ can be deduced from the {\it observed} luminosity $L_{obs}$ of each Comptonization model component and $L_s$ by the relation $L_{tot}\simeq2L_{obs}-L_s e^{-\tau}$ (see App. \ref{ampl}), where $\tau$ is the corona optical depth. This relation assumes that the corona emissions are isotropic. \\

We have plotted the values of $L_s$ and $L_{tot}$ for both coronae, as well as the amplification ratios  for the ten observations, in Fig. \ref{amplifratio}. The errors on $L_{obs}$ and $L_s$ (and consequently on $L_{tot}$) were estimated, in the case of Obs 1, from a large number of coronae model parameters (\twc, \gwc, \thc, \yhc\ and the model normalizations), taking into account their likelihood given by the contour plots shown in Fig. \ref{plotthetavstau}. The corresponding flux histograms were fitted by a Gaussian from which the 90\% error is deduced. These errors are on the order of 5\% for $L_{obs}$ and $L_s$. Because this procedure is time-consuming, we assume the same order of magnitude for the errors for all the observations. \\

While the total flux emitted by the \hot\ corona varies during the monitoring, with a significant decrease between Obs. 3 and 4,  its variation is limited to about 25\%. This is smaller than the variation in its seed soft-photons luminosity, which reaches a maximum in Obs. 5, 70\% higher than the luminosity at the beginning or the end of the monitoring. 
Concerning the \warm\ corona, $L_s$ and $L_{tot}$ vary similarly, also reaching their maximum in Obs. 5, but $\sim$35\% higher than the luminosities estimated in Obs. 1 and 10.

The {\sc warm} and {\sc hot} coronae have clearly different amplification ratios (see Fig. \ref{amplifratio}e). This is remarkably constant and equal to $\sim$2 for the {\sc warm} plasma as expected given its large optical depth (see Eq. \ref{eqratio}) and in between 5-15 for the {\sc hot} corona, with a significant decrease between Obs 3 and 6.\\

\section{Discussion}
\label{discu}

\subsection{The disk-corona geometry}
\label{corgeom}
As already said in Sect. \ref{comptmodel}, no link between the comptonised spectrum and the soft UV emission is imposed a priori in our fitting procedure.  The model simply adjusts its parameters to fit the data. We can, however, compare our best-fit values to theoretical expectations and constrain our own scenario.\\

The spectral shape of Comptonization models depends on the temperature and optical depth of the corona. Another important parameter is the geometry of the disk-corona system, which can be related to the amplification ratio. In steady states, the corona temperature adjusts, for a given optical depth, to keep this ratio to a constant value.  For instance, in the case of a plane-parallel corona geometry in thermal equilibrium above a passive disk, this amplification ratio is expected to be on the order of 3 for a thin ($\tau<1$) corona and 2 in the thick case (e.g. \citealt{haa91} and see App. \ref{ampl}). Then, photon-starved (respectively photon-fed) geometries, i.e. geometries where the corona is undercooled (respectively overcooled) compared to the plane geometry, have amplification ratios higher (respectively lower) than these theoretical values.\\

\begin{figure}
\includegraphics[width=\columnwidth,angle=0]{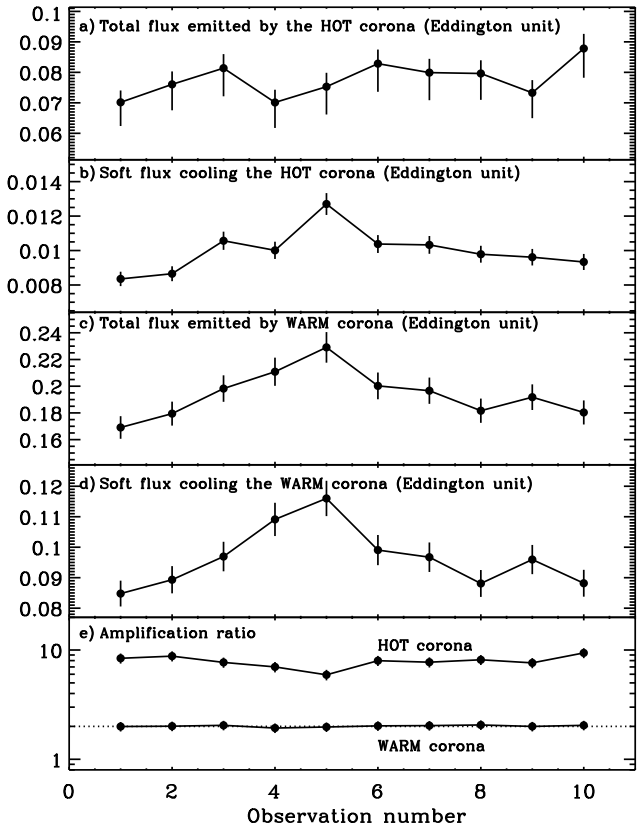}
 \caption{{\bf a)} Total flux $L_{tot}/L_{Edd}$ emitted by the \hot\ corona  and {\bf b)} {\it intercepted} soft photon luminosity $L_s/L_{Edd}$ entering and cooling the \hot\ corona.  {\bf c)} and {\bf d)}, like {\bf a)} and {\bf b)} but for the \warm\ corona.  {\bf e)} Compton amplification ratio $L_{tot}/L_s$ for the \warm\ (bottom curve) and \hot\ (top curve) coronae, respectively. The dotted line in {\bf e)} is the expected value for a thick corona in radiative equilibrium above a passive disk. All the luminosities are in Eddington units assuming a black-hole mass of 1.4$\times 10^8 M_{\sun}$ . The error bars on $L_{obs}$ and $L_s$ have been estimated as on the order of 5\%.}
\label{amplifratio}
\end{figure}

We have overplotted in Fig. \ref{plotthetavstau} the predicted relationships in steady state between the temperature and the optical depth assuming a soft-photon temperature of 10 eV and different Compton amplification ratios. 
In agreement with the estimated amplification ratios plotted in Fig. \ref{amplifratio}e, the disk-\hot\ corona system is clearly in a ``photon-starved'' configuration, with amplification ratios above 3. This suggests a disk-corona geometry where the solid angle under which the corona ``sees'' the accretion disk is small. This is a common result for Seyfert galaxies. Indeed, the slab geometry cannot reproduce hard X-ray spectra ($\Gamma_{2-10}<$2) like those observed in Seyfert galaxies because of the strong cooling that is expected from the accretion disk in such geometry (e.g. \citealt{haa93,haa94}). A photon-starved disk-corona configuration means less cooling compared to the slab geometry hence harder spectra. 

In contrast, the {\sc warm} corona agrees very well with a configuration of an optically thick plasma covering a passive disk. This is the direct consequence of a large optical depth (\tawc $\gg 1$) and our assumption to fit the Opt-UV and soft X-ray data with a unique Comptonization component (see Eq. \ref{eqratio} with no disk intrinsic emission). The slight disagreement, in Fig. \ref{plotthetavstau}, of the predicted relationship in steady state $L_{tot}/L_s= 2$ and the contours of the temperature vs. the optical depth  comes from the fact that the theoretical prediction, correct for optically thin media, is slightly incorrect at high optical depth. 

%

\subsection{The physical origin of the \warm\ and \hot\ coronae}
The sum of the seed soft-photons luminosities $L_{s, wc} + L_{s,hc}$ (dominated by $L_{s, wc}$ ) crossing both coronae  is on the order of $\sim 10^{45}$ \ergs\  for the total spectrum. Assuming that it is produced by a multicolor accretion disk around a super-massive black hole of 1.4$\times 10^8 M_{\sun}$, with an inner-disk radius of 6$R_g$ ($R_g$ the gravitational radius associated to the black hole) and an inclination angle of 30 degrees, its expected inner-disk temperature should be on the order of 15--20 eV. This is significantly more than the soft-photon temperature $\sim$3 eV of the \warm\ corona but significantly less than the soft-photon temperature $\sim$100 eV of the \hot\ corona.

 With the same reasoning, the soft-photon temperature of $\sim$3 eV of the \warm\ corona corresponds to a disk radius of $\sim$30 $R_g$. Given the expected slab geometry of the disk-\warm\ corona system, it suggests that the \warm\ corona is preferably located at this distance on the disk. On the other hand, the high temperature of the \hot\ corona and the temperature of $\sim$100 eV of its seed soft photons suggest a hot plasma in the inner region of the accretion flow. 

\subsubsection{\warm\ corona: the disk surface layer?}
%

\label{discrepro}
That we manage to fit the Opt-UV and soft X-ray data with a unique Comptonization component of large optical depth implies an amplification ratio of the \warm\ corona equal to 2 (see Eq. \ref{eqratio}), i.e., the theoretical value expected in the case of a thick corona above a passive disk. This strongly suggests that the regions of the accretion disk responsible for the optical-UV emission, which acts as the source of seed photons for the \warm\ plasma, are essentially heated by the \warm\ plasma itself. 

With an optical depth of $\sim$ 20 and a temperature of 500-600 eV, this (plane) \warm\ corona could be seen as the warm skin layer of the accretion disk. In this case the soft emission peaking at 3 eV would come from regions deeper in the disk and heated by the warm layer. Such vertical stratification of the disk is clearly expected in the case of irradiated accretion disks, the upper layer being heated to the Compton temperature of the illuminating flux (e.g. Nayakshin, Kazanas \& Kallman 2000; Nayakshin 2000; Ballantyne, Ross \& Fabian 2001b, Done \& Nayakshin 2001). \\

The optical depth of $\sim$ 20 deduced from our fits appears, however, to be well above the expected theoretical values, on the order of a few units. This could be explained by the very simple assumptions of our corona models, which assume a constant temperature and optical depth for the whole corona. The inclination of the accretion disk could also play a role, since it would increase the effective optical depth compared to the theoretical estimates, which are computed in the vertical direction of the layer. \\

A natural origin of the illuminating radiation producing the \warm\ corona could be the \hot\ corona itself (see next section). However, the \hot\ corona luminosity is about a factor 2 lower than the total \warm\ corona luminosity (see Fig. \ref{amplifratio}); so irradiation is only part of the heating mechanism of the upper layers of the accretion disk forming the \warm\ corona. Besides, the amplification ratio of 2 disfavors intrinsic emission produced by the disk deeper layers but instead suggests that the \warm\ corona has its own heating process.
If this interpretation is correct, it means that the accretion heating power is most likely to be liberated at this disk surface layer and not in the disk interior. Interestingly, recent 3-D radiation-MHD calculations, when applied to protostellar disks, show that the disk surface layer can indeed be heated by dissipating the disk magnetic field, while the disk interior is ``magnetically dead'' and colder than a standard accretion disk \citep{hir11}. The same processes could occur in the disk of Mrk 509, giving birth to the \warm\ corona needed in our fits.\\

It is worth noting, however, that these conclusions are model-dependent, and for instance, they strongly depend on our assumption to fit the optical-UV and soft X-ray emission with a unique Comptonization component. The addition of a pure multicolor-disk component in the optical-UV band would necessarily decrease the importance of the \warm\ corona emission in this energy range and consequently would decrease its amplification ratio. From a statistical point of view, there is no need to add such a multicolor disk component, and thus, with this limit in mind, we can conclude that the \warm\ corona likely covers and heats a large part of the accretion disk.\\

\subsubsection{\hot\ corona: the hot inner flow?}
\begin{figure}
\includegraphics[height=1.1\columnwidth,angle=-90]{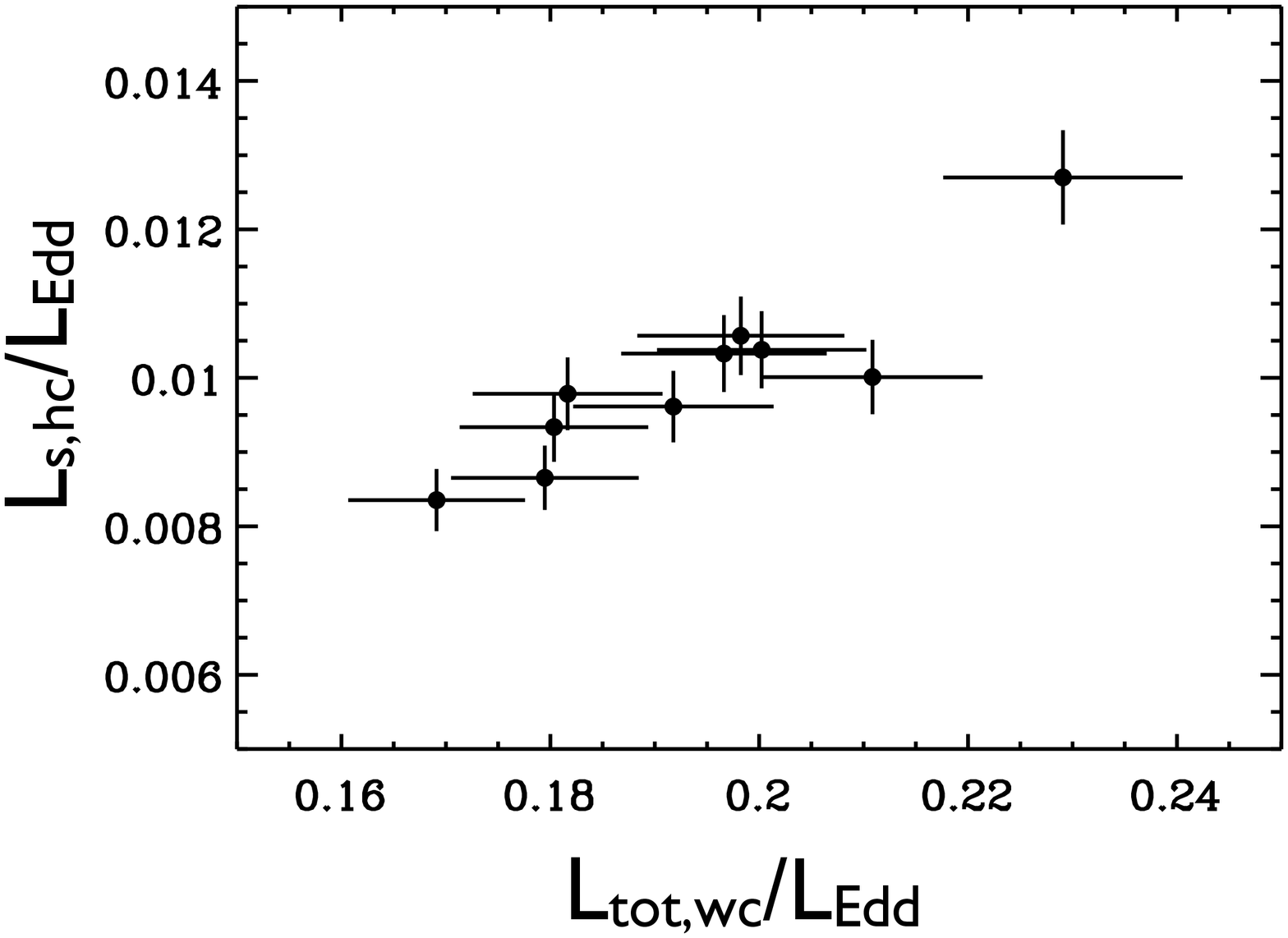} 
\vspace*{-0.5cm}\caption{Correlation of the seed soft-photon luminosity of the \hot\ corona, $L_{s,hc}$, and the total luminosity of the \warm\ plasma $L_{tot,wc}$.}
\label{lscompltotnth}
\end{figure}
\begin{figure*}
\begin{center}
\includegraphics[width=0.9\textwidth,angle=0]{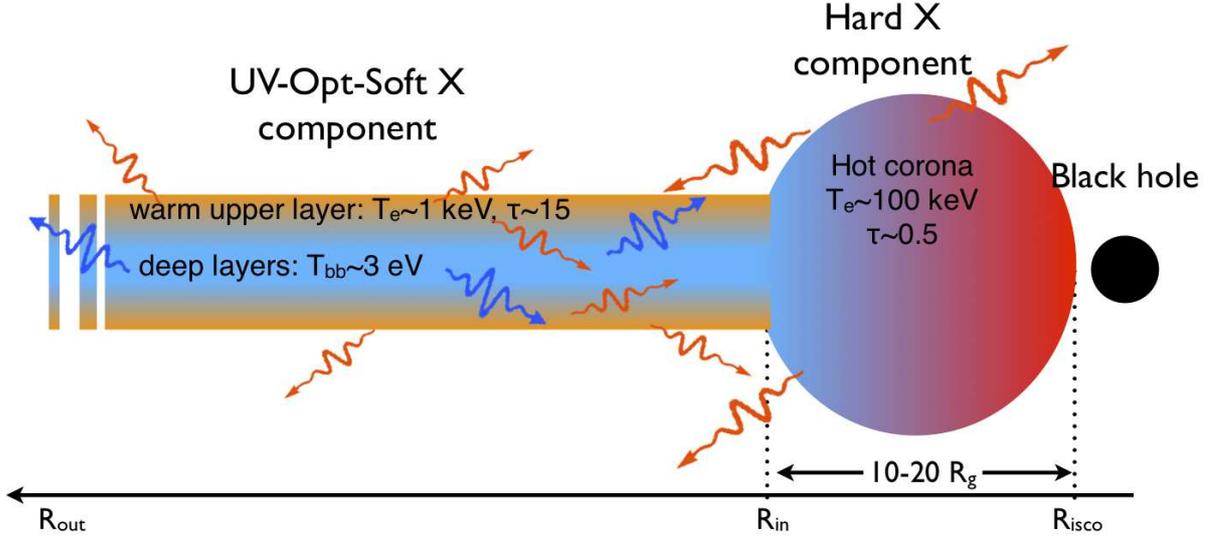} 
 \caption{A sketch of the accretion flow geometry in the inner region of the galaxy Mrk 509. The \hot\ corona, producing the hard X-ray emission, has a temperature of $\sim$ 100 keV and optical depth $\sim$ 0.5. It is localized in the inner part of the accretion flow ($R<R_{in}$) and illuminates the accretion disk beyond $R_{in}$, helping to form a \warm\ layer at the disk surface. This \warm\ component has a temperature of $\sim$ 1 keV and an optical depth $\sim$ 15 and produces the optical-UV up to soft X-ray emission. It  extends over a large part of the accretion flow, heating the deeper layers and comptonizing their optical-UV emission. In return, part of this emission enters and cools the \hot\ corona.}
\label{sketch}
\end{center}
\end{figure*}
The total power emitted by the \hot\ corona is on the order of 1.3$\times 10^{45}$ \ergs\ i.e. $\sim$5\% of the Eddington luminosity of a supermassive black hole of 1.4$\times 10^{8} M_{\sun}$. There are a few theoretical models that explain the presence of such hot thermal plasma at such high luminosity: for instance the luminous hot accretion flow (LHAF)  proposed by \citealt{yua01} which is an extension of the ADAF solutions to high accretion rate (see also \citealt{yua06}). Jet emitting disks (JED) solutions \citep{fer06a} also provide a hot and optically thin plasma. Since no powerful and collimated jet is observed in Mrk 509, the JED solution is apparently not appropriate. Aborted jets have been proposed, however, to explain the properties of Seyfert galaxies \citep{hen97,pet97,ghi04}. A JED could then be present but without filling all the conditions for producing a ``real'' jet, such as its insufficiently large radial extension. The \hot\ corona could be present but not the jet.\\

The 100 eV temperature for the soft seed photons of the \hot\ corona is a factor $\sim$5 above the value expected from a standard accretion disk. A few physical arguments could help to solve this inconsistency. First, the disk spectra are certainly more complex than the simple sum of blackbodies due to the possible effects of electron scattering in comparison to absorption. Special and general relativistic effects could also play a significant role on the observed emission. These effects lead to modified black-body spectra with an "observed" temperature higher than the effective temperature by a factor between 2 and 3 (e.g. \citealt{cze87,ros92,mer00}). The presence of the \warm\ corona above a large part of the disk can also strongly modify the disk emission, as shown in our fits \citep{cze87,cze03,don12}. 

An interesting interpretation could be that the seed soft photons of the \hot\ corona are those emitted by the \warm\ plasma. The \warm\ plasma emission peaks at a few eV and extends to a few 100's of eV, so it can be ``seen'' by the \hot\ corona as a soft photon field with an intermediate temperature of $\sim$ 100 eV. Such an interpretation cannot be modeled easily with \xspec, but it is nicely supported by the strong correlation (with a linear Pearson correlation coefficient equal to 0.92, which corresponds to a $>$99\% confidence level for its significance) between the seed soft photon luminosity of the \hot\ corona, $L_{s,hc}$, and the total luminosity of the \warm\ plasma $L_{tot,wc}$ (see Fig. \ref{lscompltotnth}). 

\subsubsection{A tentative toy picture}
\label{toypicture}
From the above discussions, a tentative toy picture of the overall configuration of the very inner regions of Mrk 509 could be the following: a \hot\ corona ($T_e\sim$ 100 keV, $\tau$=0.5) fills the inner part of the accretion flow, below say $R_{in}$. It illuminates the outer part of the disk and contributes to the formation of a warm upper layer ($T_e\sim$ 1 keV, $\tau$=20) on the disk beyond $R_{in}$, the \warm\ corona. This \warm\ corona also possesses its own heating process, potentially through dissipation of magnetic field at the disk surface. It  heats the deeper layers of the accretion disk, which then radiates like a black body (at $T_{bb}\sim$3 eV). This \warm\ corona upscatters these optical-UV photons up to the soft X-ray, a small part of them entering and cooling the \hot\ corona. The small solid angle under which the hot flow sees the warm plasma agrees with the large amplification ratio of $\sim$5--10 estimated from our fits.  The 100 eV temperature obtained by our \xspec\ modeling would be explained by the fact that \xspec\ would try to mimic the \warm\ plasma emission, which covers an energy range between $\sim$ 1 eV and $\sim$ 1 keV, with the multicolor-disk spectral energy distribution of the seed soft photons available in \compps. A tentative sketch is shown in Fig. \ref{sketch}.  \\

Clearly, in this picture the two coronae should be powered by the same accretion flow and should consequently be energetically coupled. Assuming that the \warm\ corona covers the accretion disk between $R_{out}$ and $R_{in}$ and then that the \hot\ corona fills the space between $R_{in}$ and $R_{isco}$, then the accretion power liberated in the \warm\ corona is
\begin{equation}
P_{acc,wc}=\frac{GM\dot{M}}{2R_{in}}\left [1-\frac{R_{in}}{R_{out}}\right ]
\label{eq1}
\end{equation}
while the accretion power liberated in the \hot\ corona is
\begin{equation}
P_{acc,hc}=\frac{GM\dot{M}}{2R_{isco}}\left [1-\frac{R_{isco}}{R_{in}}\right ],
\label{eq2}
\end{equation}
so the ratio between the two is
\begin{equation}
\frac{P_{acc,hc}}{P_{acc,wc}}=\frac{R_{in}}{R_{isco}}\frac{1-\frac{R_{isco}}{R_{in}}}{1-\frac{R_{in}}{R_{out}}}.
\label{ratio}
\end{equation}
Neglecting the advection, which seems reasonable given the accretion rate of the system, the accretion power liberated in each corona is close to their total luminosity. The ratio $P_{acc,hc}/P_{acc,wc}$ is then on the order of $L_{tot,hc}/L_{tot,wc}$ and Eq. \ref{ratio} gives a one-to-one relationship between the total luminosities $L_{tot,hc}$ and $L_{tot,wc}$  (or, equivalently, $L_{tot,hc}$ and $L_{tot,hc}/L_{tot,wc}$) and the inner disk radius $R_{in}$ and total accretion rate $\dot{M}$, as soon as $R_{isco}$ and $R_{out}$ are given. \\

\begin{figure}
\hspace*{-1cm}\includegraphics[height=1.3\columnwidth,angle=-90]{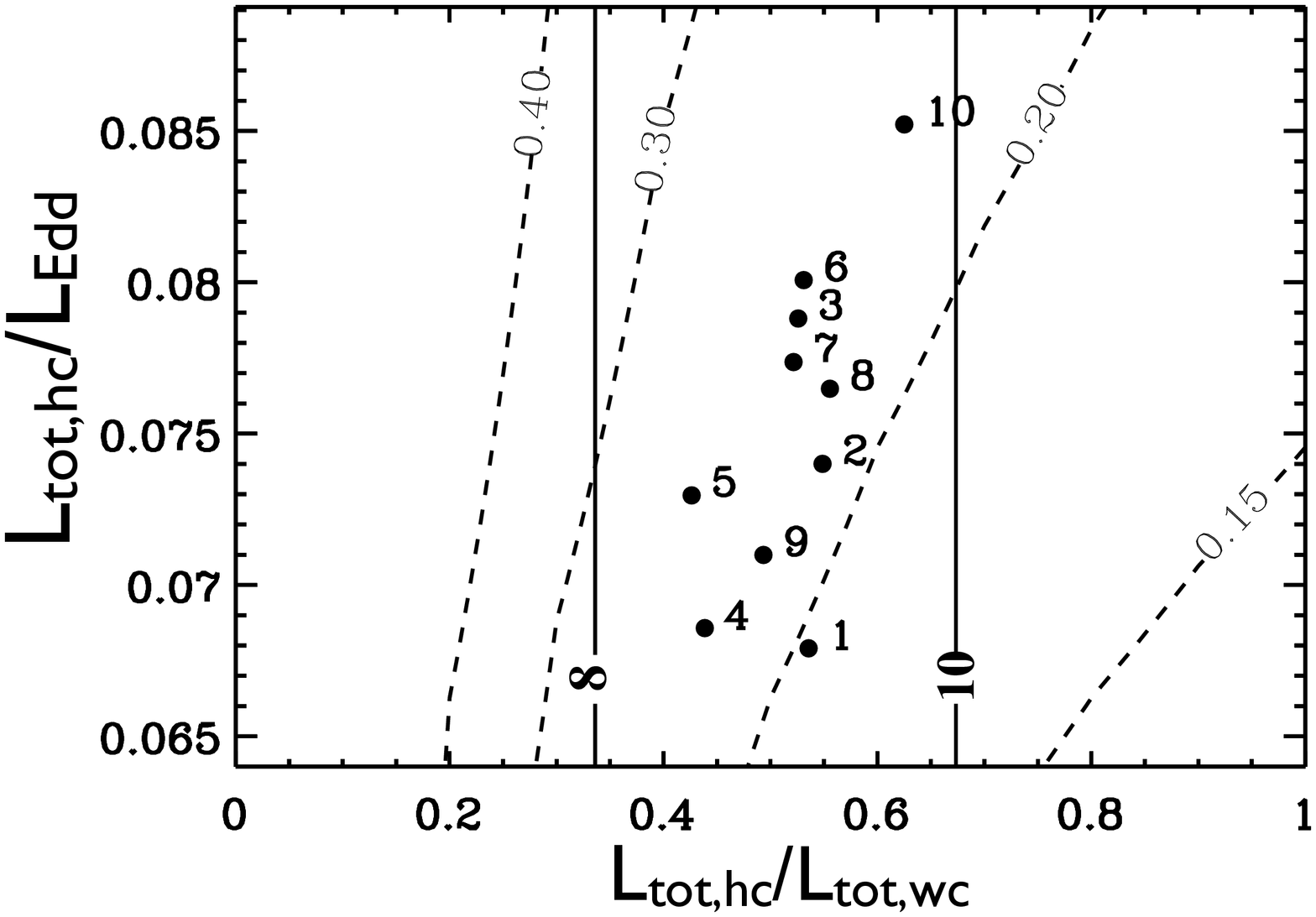} 
\vspace*{-0.5cm}\caption{Contour plots of the inner disk radius $R_{in}$ (solid lines, in units of $R_g$) and the total accretion rate $\dot{M}$ (dashed lines, in Eddington units and assuming an accretion efficiency $\eta=(2R_{isco}/R_g)^{-1}$) in the $L_{tot,hc}$--$L_{tot,hc}/L_{tot,wc}$ plane (see Eqs. \ref{eq1} and \ref{eq2}). The points indicate the positions of the 10 observations.}
\label{ltotrinmdot}
\end{figure}

We have mapped, in Fig. \ref{ltotrinmdot}, the $R_{in}$--$\dot{M}$ plane into the $L_{tot,hc}$--$L_{tot,hc}/L_{tot,wc}$ plane assuming $R_{isco}$=6$R_g$ and $R_{out}$=100$R_g$. The observed total luminosities agree with Eq. \ref{eq1} and Eq. \ref{eq2} if $R_{in}$ is between 8 and 10 $R_g$ and the accretion rate $\dot{M}$ between 20 and 30\% of the Eddington rate (assuming an accretion efficiency $\eta=(2R_{isco}/R_g)^{-1}$). 
$R_{in}$ strongly depends on $R_{isco}$ and decreases as $R_{isco}$ decreases. A value of 10$R_g$ can then be seen as a rough upper limit. On the other hand, $\dot{M}$ only slightly depends on $R_{isco}$. Both $R_{in}$ and $\dot{M}$ very slightly depend on $R_{out}$ unless $R_{out}$ become very small ($<$ 10-20 $R_g$).\\

\subsection{A luminosity-independent spectral shape for the soft X-ray excess}
That the \warm\ corona, which is at the origin of the Opt-UV to soft X-ray emission in our modeling of Mkr 509, has naturally a constant amplification ratio  has an interesting consequence. It implies a spectral emission that is almost independent of the AGN luminosity. Indeed, the photon index of the comptonized spectrum is directly linked to the amplification ratio (e.g. \citealt{bel99,mal01}). A constant amplification ratio then implies a roughly constant spectral index. { This also depends on the soft photon temperature, but this effect is weak as far as this temperature is not varying too much}. Moreover, in radiative equilibrium the corona temperature and optical depth follow a univocal relationship. For a given optical depth $\tau$ corresponds a unique temperature $kT$ and, to the extent that the radiative equilibrium is insured, there is no reason for $\tau$ and $kT$ to vary. The temperature and optical depth of the \warm\ corona are indeed consistent with a constant during the monitoring (see Fig. \ref{lcparam}).  Thus, in these conditions, we expect an almost constant spectral shape for the soft X-ray excess, regardless  of the luminosity of the AGN, in very good agreement with observations (e.g. \citealt{gie04,cru06,pon06}).

\subsection{Geometry variations}
 Simulations by Malzac \& Jourdain (\citeyear{mal01}) show that the time scale for a disk-corona system to reach equilibrium, after
  rapid radiative perturbation in one of the two phases, is of a few corona light crossing times i.e. shorter than a day. We
  can thus (reasonably) assume that each of the ten spectra of Mrk 509 correspond to a disk-corona
  system where hot and cold phases are in radiative balance with each other.  In this case, a fixed
  disk-corona configuration in radiative equilibrium should correspond to
  a constant heating/cooling ratio. As already said, this is what we observe in the case of the \warm\ corona, its amplification ratio being necessarily (given our model) equal to 2 (see Fig. \ref{amplifratio}e).  In the case of the \hot\ corona, however, the amplification ratio decreases in the middle of the campaign (Obs 4 and 5) suggesting a variation in the disk-\hot\ corona configuration during the monitoring.\\ 

A decrease in the amplification ratio means that the disk-\hot\ corona geometry becomes less photon-starved. During Obs 4 and 5 the  total luminosity of the \warm\ corona increases (cf. Fig. \ref{amplifratio}c), while the total luminosity of the \hot\ corona decreases (cf. Fig. \ref{amplifratio}a). If the above picture is correct, these variations should be linked somehow to variations in the inner accretion disk radius $R_{in}$ and/or of the total accretion rate.  Looking to Fig. \ref{ltotrinmdot}, variations of $\sim$5-10\%  for these parameters are enough to explain the observed luminosity variations. The variation in $R_{in}$ can also explain the variation in the \hot\ corona amplification factor.\\

\subsection{A pair-dominated hot corona?}
\label{paircorona}
From the luminosities $L_{tot}$ and $L_s$, we can estimate the compactness $l_{tot}=L_{tot}\sigma_T/(R m_e c^3)$ and $l_{s}=L_{s}\sigma_T/(R m_e c^3)$ where $\sigma_T$ is the Thomson cross section, $m_e$ the electron mass, $c$ the speed of light, and $R$ the typical size of the corona. We assume $R=10R_g$. Then, on average for the 10 observations, the total compactness $l_{tot,hc}\sim$100, and the soft compactness $l_{s,hc}\sim$ 15 for the \hot\ corona, while $l_{tot,wc}\sim$230 and $l_{s,wc}\sim$115 for the \warm\ corona.
For such values of $l_{tot}$ and $l_s$ (and thus $l_h=l_{tot}-l_s$ the heating compactness), no pair equilibrium can be reached for the \warm\ corona given its very low temperature (cf. \citealt{ghi94}, G94 hereafter, and their Fig. 2). This is different for the \hot\ corona, which has a high enough temperature, given its values of $l_{tot}$ and $l_s$, to be dominated by pairs. \\
\begin{figure}
\includegraphics[width=\columnwidth,angle=0]{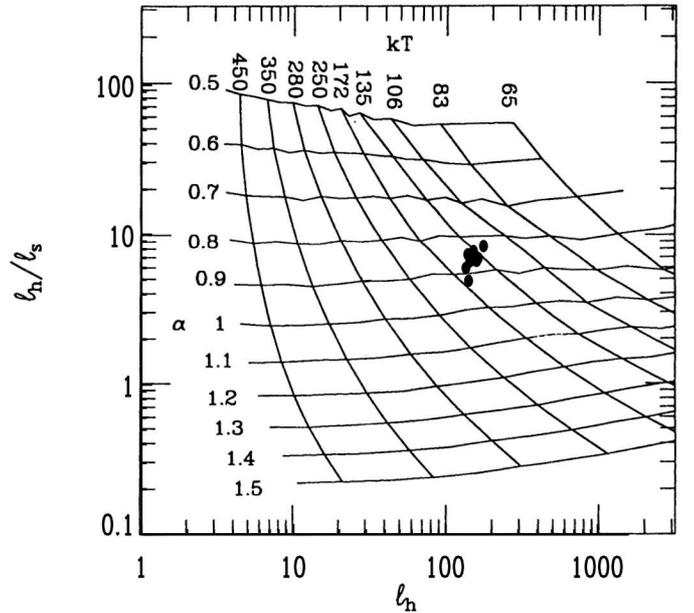} 
 \caption{Hard-to-soft compactness ratio $l_h/l_s$ vs. hard compactness $l_s$. The figure is extracted from \cite{ghi94}. It shows the one-to-one correspondence between the spectral characteristics  of the Compton spectrum of a pure pair plasma (energy spectral index $\alpha$ and temperature $kT$) and the soft and hard compactnesses.The black points correspond to the 10 observations of Mrk 509.}
\label{pairsfig}
\end{figure}

For a pure pair plasma, there is a one-to-one relationship between the hard and soft compactnesses and the spectral characteristics (energy spectral index $\alpha$ and temperature $kT$) of the Compton spectrum. G94 have mapped the plane $\alpha$-$kT$ into the plane ($l_h/l_s$-$l_h$). We have reported their Fig. 2 in our Fig. \ref{pairsfig}, overplotting the values of $l_{h,hc}$ and $l_{s,hc}$  computed for the 10  observations of Mrk 509. 

Astonishingly enough, given the simple assumptions of the G94 model (homogeneous spherical corona in a homogeneous black body photon field) and our approximated estimates of the compactnesses, they agree nicely with the theoretical expectations of a pure pair plasma. The observed spectral index  $\alpha\sim$0.7-0.8 (see Table 1, with $\alpha=\Gamma-1)$, the \hot\ corona temperature on the order of 100 keV, and the variations of these parameters during the monitoring (i.e. $\Delta\alpha\sim 0.1$ and $\Delta kT\sim$ 80 keV) are not far from the expected ranges. {While this comparison is admittedly rough (e.g. the observed spectral index $\alpha$ should have been steeper between 0.8 and 0.9 to completely agree with the expectation), it is interesting to notice that the \hot\ corona behaves spectrally as if it is dominated by pairs}.\\

In the case of a pair-dominated plasma, the increase in the cooling, for a constant heating, is expected to produce an increase in the plasma temperature and not a decrease. This can be seen in Fig. \ref{pairsfig}. Fixing $l_h$ but increasing $l_s$  shifts the pair plasma to a higher temperature state. This can be understood as follows (see G94): increasing the cooling first decreases the plasma temperature. But this implies a decrease in the pair creation rate, i.e., a decrease in the number of particles. The available energy is thus shared among fewer particles and eventually, when pair equilibrium is reached again, the temperature is higher. 

From our fits, the soft compactness $l_{s,hc}$ of the \hot\ corona varies by  $\sim$60\% during the monitoring, while the heating compactness $l_{h,hc}$ varies by $\sim$15\%. Thus we are roughly in the case of a constant heating with variable cooling. Interestingly enough, the \hot\ corona temperature deduced from our modeling (marginally) correlates with  $l_{s,hc}$ as expected for a pair-dominated plasma. The linear Pearson correlation coefficient is equal to 0.67, which corresponds to a 95\% confidence for its significance.

%
\begin{figure*}
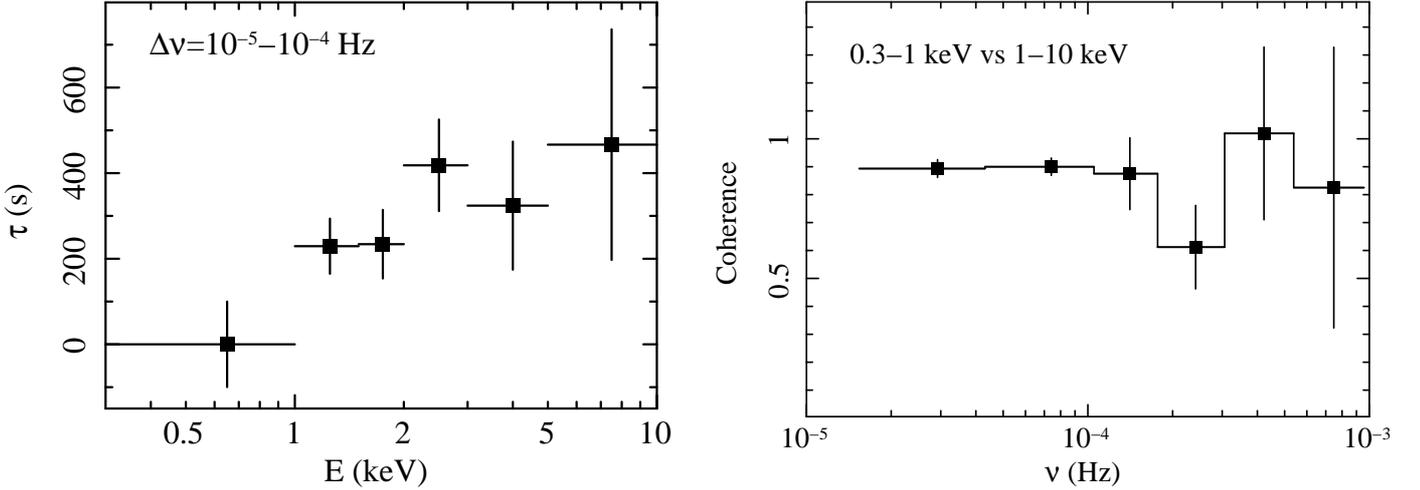

\begin{tabular}{cc}
\includegraphics[height=\columnwidth,angle=-90]{tlE_lowfreq_NEW.ps} &\includegraphics[height=1.02\columnwidth,angle=-90]{coher.ps}
\end{tabular}
 \caption{ {\bf Left:} Lag energy spectrum. It is obtained by computing time lags in Fourier frequency space, between the energy band 0.3-10 keV and a number of adjacent energy channels, averaging the resulting lags over the frequency interval 10$^{-5}$-10$^{-4}$ Hz. {\bf Right:} Coherence between 0.3-1 and 1-10 keV energy bands as a function of Fourier frequency once a correction for the Poisson noise effects is applied \citep{vau97}.\label{timelagfig}}
\end{figure*}

\subsection{Some hint of the behavior on short time scale?}
To investigate the short time scale correlations between spectral components in the X-ray band, we estimated the coherence function (e.g.  \citealt{vau97})
between the 0.3-1 keV and 1-10 keV energy bands by making use of the XMM-Newton data (see right plot of Fig. \ref{timelagfig}). 
The two energy bands display a high level of linear correlation over all the sampled frequencies (corresponding to time scales between 60 and 0.5 ks).
This behavior is at odds with the observed lack of correlated variability on longer time scales (see Fig. \ref{corr} and Sect. \ref{lc}). However, this result is consistent with the existence of a small and compact \hot\ corona, which is able to produce coherent variations on short time scales.\\

We explored the causal connection between the different spectral
components by computing time lags as a function of energy (see left plot
of Fig. \ref{timelagfig}). The time lags are {first} computed in the { Fourier} frequency
domain between the entire 0.3-10 keV { reference} band and adjacent channels,
with the channel being subtracted from the reference at each step, so as
to remove the contribution from correlated Poisson noise (e.g. \citealt{zog11}). { The derived lags are then averaged over the range of frequencies  $\Delta\nu\sim10^{-5}-10^{-4}$ Hz, i.e., the frequency domain not dominated by Poisson noise, and plotted as a function of energy.} 
The lag spectrum has been shifted so that the zero-lag energy channel corresponds to the leading one. 
Hard positive lags (variations in the hard channels being delayed with respect to those in the soft channels) are observed over the full energy band, the relative amplitude of the measured lags increasing with the energy separation between the channels. This trend is commonly observed in black hole X-ray binaries, as well as in some AGNs in the low-frequency range (e.g. \citealt{now99,pap01,are06,zog11,utt11,demar12}). 


These hard positive lags are generally interpreted by the propagation of mass accretion rate fluctuations in the disk (e.g. \citealt{kot01,are06}) and, in order to be observed in the soft X-ray bands, the soft X-ray emitting region has to be strongly connected somehow to the disk accretion flow. This is naturally expected if the soft X-ray emitting region is the upper layer of the accretion disk, as we propose for the \warm\ corona.

\subsection{Do we have all the pieces of the puzzle?}
Most of our results are model-dependent. For instance, we could add other spectral components (like e.g. a blurred reflection or a pure disk black body) to fit the UV and soft X-rays. The possibilities are numerous and may give a different picture of the inner regions of Mrk 509 than the one revealed by our modeling.
Our choice was instead to choose a small number of spectral components. Our main assumption was that the optical-UV emission, observed in the optical monitor of \xmm, was the signature of an optically thick multi-color accretion disk whose optical-UV photons are comptonised in two different media: the first one (the \warm\ corona) producing the soft X-ray excess and the second one (the \hot\ corona) the high-energy continuum emission. In this framework it appears that
\begin{itemize}
\item The \warm\ corona has an amplification ratio equal to two i.e. the expected value of an optically thick plasma above a passive accretion disk, the disk emission being dominated by the reprocessing. In consequence, most of the disk emission is due to the heating from the \warm\ corona itself. The \warm\ corona could be the warm upper layer of the accretion disk. 
\item The \hot\ corona amplification ratio is between five and ten and thus suggests a more photon-starved geometry. The temperature of the soft seed photons entering and cooling the \hot\ corona, on the order of 100 eV, suggests a location in the inner region of the accretion flow.
\item A 100 eV  temperature is about a factor 5 above the expected value for a standard accretion disk around a 1.4$\times 10^8 M_{\sun}$ black hole. Effects of electron scattering and of special and general relativity could significantly affect the observed emission and lead to modified black body spectra with an ``observed'' temperature higher than the effective temperature. The seed soft photons of the \hot\ corona can also be part of the \warm\ corona emission. The 100 eV temperature obtained by our \xspec\ modeling would be explained by the fact that \xspec\ would try to mimic the \warm\ plasma emission, which covers an energy range between 3 and $\sim$ 1 keV, with the multicolor-disk spectral energy distribution available in \compps\ for the seed soft-photon field.
\item The \hot\ corona may be dominated by pairs.
\item The observed flux and spectral variability are mainly due to changes in coronae geometries.
\end{itemize} 


The toy picture presented in Sect. \ref{toypicture} broadly agrees with the non-detection of a relativistically broadened iron line in Mrk 509. Indeed, P12 found that the resolved component of the iron line profile ($\sigma$=0.22 keV) agrees with an origin beyond 40$R_g$ from the black hole. This supports the absence of a neutral optically thick accretion disk down to $R_{isco}$. On the other hand, a weak ionized iron line component is present and can be modeled with a relativistic line produced, in an ionized disk, down to $\sim$10 $R_g$ from the black hole.  The presence of this ionized matter agrees with a  \warm\ upper layer on the accretion disk.


It has been recently suggested by \cite{don12} that the accretion disk in AGNs could evolve radially from an optically thick accretion disk beyond a transition radius called $R_{corona}$ to a two-phase accretion flow at smaller radii, formed by an optically thick Comptonized disk and an optically thin corona.  The latter would play the role of our \hot\ corona, while the optically thick Comptonized disk will produce the \warm\ Comptonized emission.  The energetic coupling between the two coronae is taken into account in this model. It possesses the different components used in our fits (the accretion disk and the \warm\ and \hot\ plasmas). However, it assumes that the \warm\ and \hot\ coronae are localized in the same region of the accretion flow, below  $R_{corona}$, with the \warm\ corona embedded in the \hot\ one (see Fig. 5 of \citealt{don12}).  This picture does not agree completely with our estimates of the amplification ratios; however, it is interesting to see that our conclusions are in broad agreement with the assumptions proposed by these authors. It will be worth testing their model on the Mrk 509 data\footnote{Their model is called {\sc{optxagn}}, and is publicly available as a local model for {\sc{XSPEC}}}. This is devoted to a future work. \\

\section{Conclusions}
\label{conc}
Mrk 509 was monitored simultaneously with \xmm\ and INTEGRAL for one month in late 2009, with one observation every four days for a total of ten observations. A series of papers has already been published using these data focusing on different aspects of the campaign. Here we analyzed the broad-band optical/UV/X-ray/gamma-ray continuum of the source during this monitoring. Our main effort was to adopt realistic Comptonization models to fit the primary continuum, the accretion disk, and the soft X-ray excess.\\

We obtained a relatively good fit to the data by assuming the existence of two clearly different media: a  hot ($kT \sim$ 100 keV), optically-thin ($\tau \sim$ 0.5) corona for the primary continuum emission, and a warm ($kT \sim$ 1 keV), optically-thick ($\tau \sim$ 10-20) plasma for the optical-UV to soft X-ray emission.\\

These two media have very different geometries: the \warm\ corona covers a large part of the accretion disk, the disk radiation being dominated by the reprocessing of the \warm\ corona emission. The \warm\ corona could be the warm upper layer of the accretion disk. The temperature of the soft-photon field cooling it is on the order of 3 eV. 

In contrast, the \hot\ corona has a more photon-starved geometry and can be the inner part of the accretion flow.  The transition radius $R_{in}$ between the outer disk/\warm\ plasma and the \hot\ inner corona is estimated to be around 10--20 $R_g$. The variation in $\dot{M}$  but also in $R_{in}$ would be at the origin of the variation in the coronae luminosities, the variation in $R_{in}$ also producing the variation in the  \hot\ corona  amplification ratio. The spectral and flux variation in the \hot\ corona are also consistent with a pair-dominated plasma.\\

These different conclusions may depend at some level on the choice of our modelling. For example, we do not include blurred ionized reflection, while part of the soft X-ray excess could be produced by this component. The presence of an (admittedly weak) ionized iron line, which agrees with a relativistic profile, by P12 would also support the presence of such reflection. On  the typical time scale ($\sim$ day) of this campaign, however, it does not seem to play a major role and Comptonization in a \warm\ plasma, covering the accretion disk, appears the most natural explanation for the soft X-ray excess in Mrk 509.

\section*{Acknowledgments}

POP acknowledges the very interesting discussions with J. Ferreira and G. Lesur . This work is based on observations obtained with \xmm, an ESA science mission with instruments and contributions directly funded by ESA Member States and the USA (NASA). It is also based on observations with INTEGRAL, an ESA project with instrument and science data center funded by ESA member states (especially the PI countries: Denmark, France, Germany, Italy, Switzerland, Spain), Czech Republic, and Poland and with the participation of Russia and the USA. This work made use of data supplied by the UK Swift Science Data Centre at the University if Leicester. SRON is supported financially by NWO, the Netherlands Organization for Scientific Research. J.S. Kaastra thanks the PI of Swift, Neil Gehrels, for approving the TOO observations, and the duty scientists at the William Herschel Telescope for performing the service observations. P.-O. Petrucci acknowledges financial support from the CNES and the French GDR PCHE. M. Cappi, M. Dadina, S. Bianchi, and G. Ponti acknowledge financial support from contract ASI-INAF n. I/088/06/0. N. Arav and G. Kriss gratefully acknowledge support from NASA/\xmm\ Guest Investigator grant NNX09AR01G. Support for HST Program number 12022 was provided by NASA through grants from the Space Telescope Science Institute, which is operated by the Association of Universities for Research in Astronomy, Inc., under NASA contract NAS5-26555. E. Behar was supported by a grant from the ISF. A. Blustin acknowledges the support of an STFC Postdoctoral Fellowship. P. Lubi\'nski has been supported by the Polish MNiSW grants N N203 581240 and 362/1/N-INTEGRAL/2008/09/0. M. Mehdipour acknowledges the support of a PhD studentship awarded by the UK Science \& Technology Facilities Council (STFC). G. Ponti acknowledges support via an EU Marie Curie Intra-European Fellowship under contract no. FP7-PEOPLE- 2009-IEF-254279. K. Steenbrugge acknowledges the support of Comit\'e Mixto ESO - Gobierno de Chile.

\begin{appendix}

\section{Optical depth estimates}
\label{optdepth}
The \nthcomp\ model used to fit the optical-UV-soft X-ray emission does not directly provide the value of the optical depth of the plasma for which it models the emission. However, the plasma optical depth  has more physical meaning than the photon index of the emitted spectrum, so it is an important parameter that we need to estimate. Generally, one compares the \nthcomp\ spectral shape, for a given set of photon index and temperature, to another model that provides the optical depth. 
While commonly used, the procedure is generally not precisely discussed. The accuracy of this comparison is important, because inaccurate values for the temperature and optical depth will give a different interpretation for the heating/cooling ratios (see Sect. \ref{corgeom}). \\

We compared the results obtained with \nthcomp\ with those from three other \xspec\ Comptonization models: \compst, \comptt, and \eqpair. \compst\  \citep{sun80} only has the corona temperature and optical depth as free parameters (no parameter for the soft photon temperature), but the range of validity is relatively wide, including low temperature and large optical depth as expected in our case. \comptt\ \citep{tit94} is a commonly used Comptonization code that provides the optical depth and the temperature. Like \compst\ it assumes a black body distribution for the soft photon field. This is the model used by M11.  \eqpair\ is a more complete Comptonization model that carefully takes most of the microphysics needed to compute realistic Comptonization spectra into account \citep{cop91}. The soft photon distribution can be either a simple black body or a multicolor disk spectrum. It can treat hybrid thermal/non-thermal particle distributions. Its natural parameters are the optical depth and the heating/cooling ratio. The temperature is an output of the code. To obtain the temperature in \xspec, the {\it chatter} command has to be larger than 15. However, for most of these models, the values expected for the temperature and optical depth of the {\sc warm} corona are outside the nominal range of validity indicated in \xspec. The temperature has to be between 2 and 500 keV for {\comptt}, between 1 keV and 1 MeV for {\nthcomp}, and the optical depth should be below 10 for \eqpair. It is thus interesting to see how these models compare. \\

\begin{figure*}
\begin{tabular}{cc}
\includegraphics[width=\columnwidth]{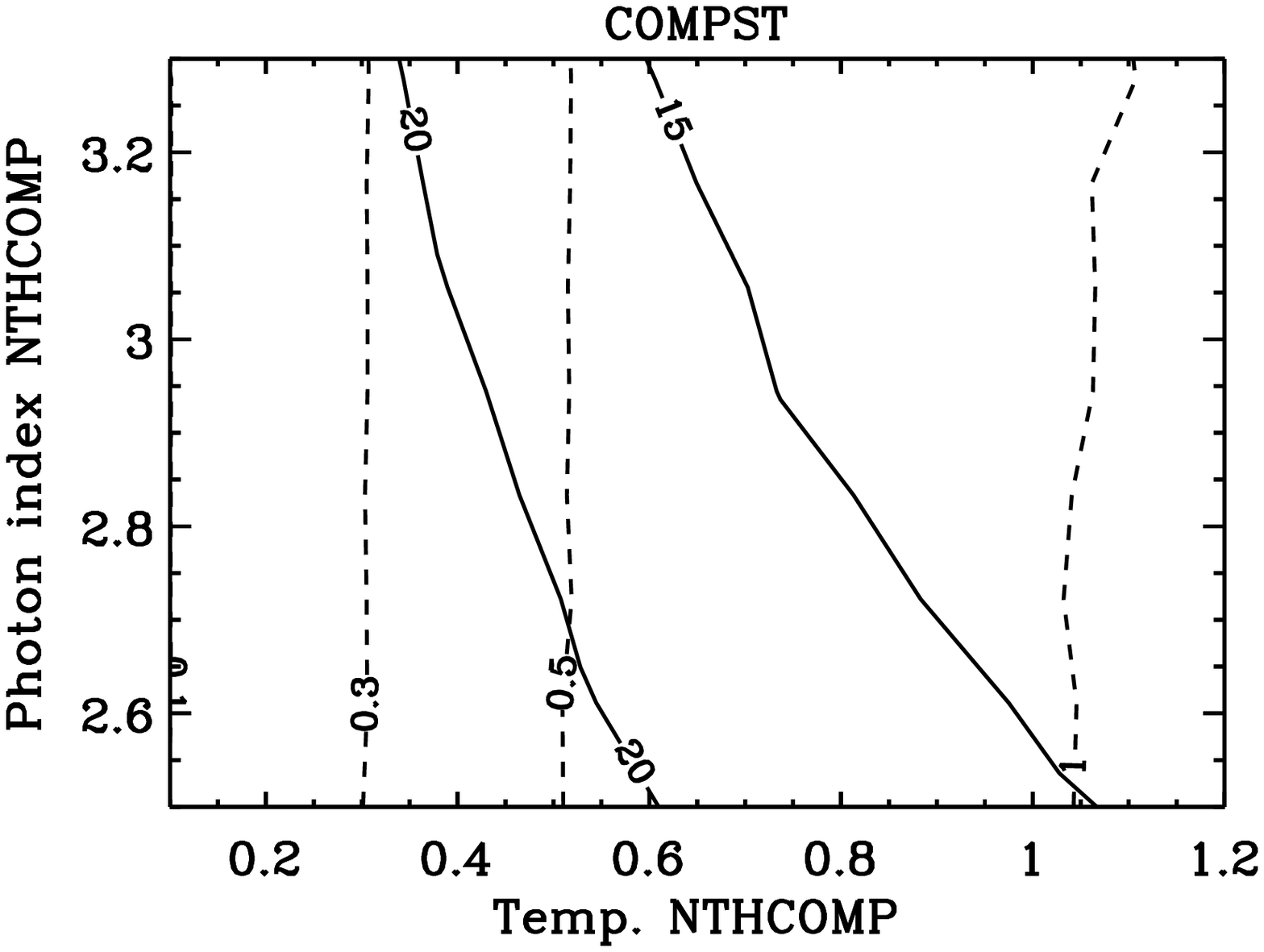}&
\includegraphics[width=\columnwidth]{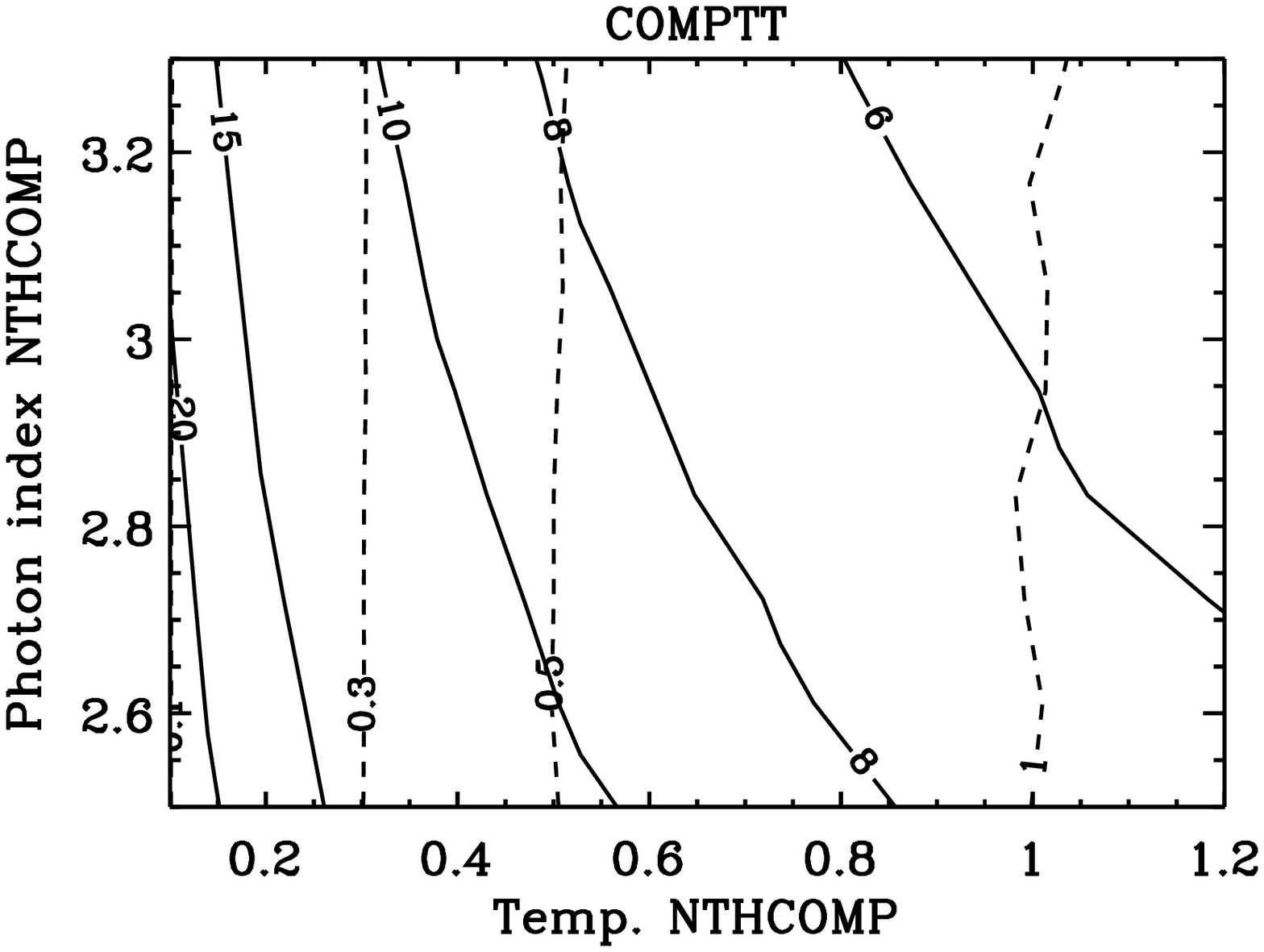}\\
\includegraphics[width=\columnwidth]{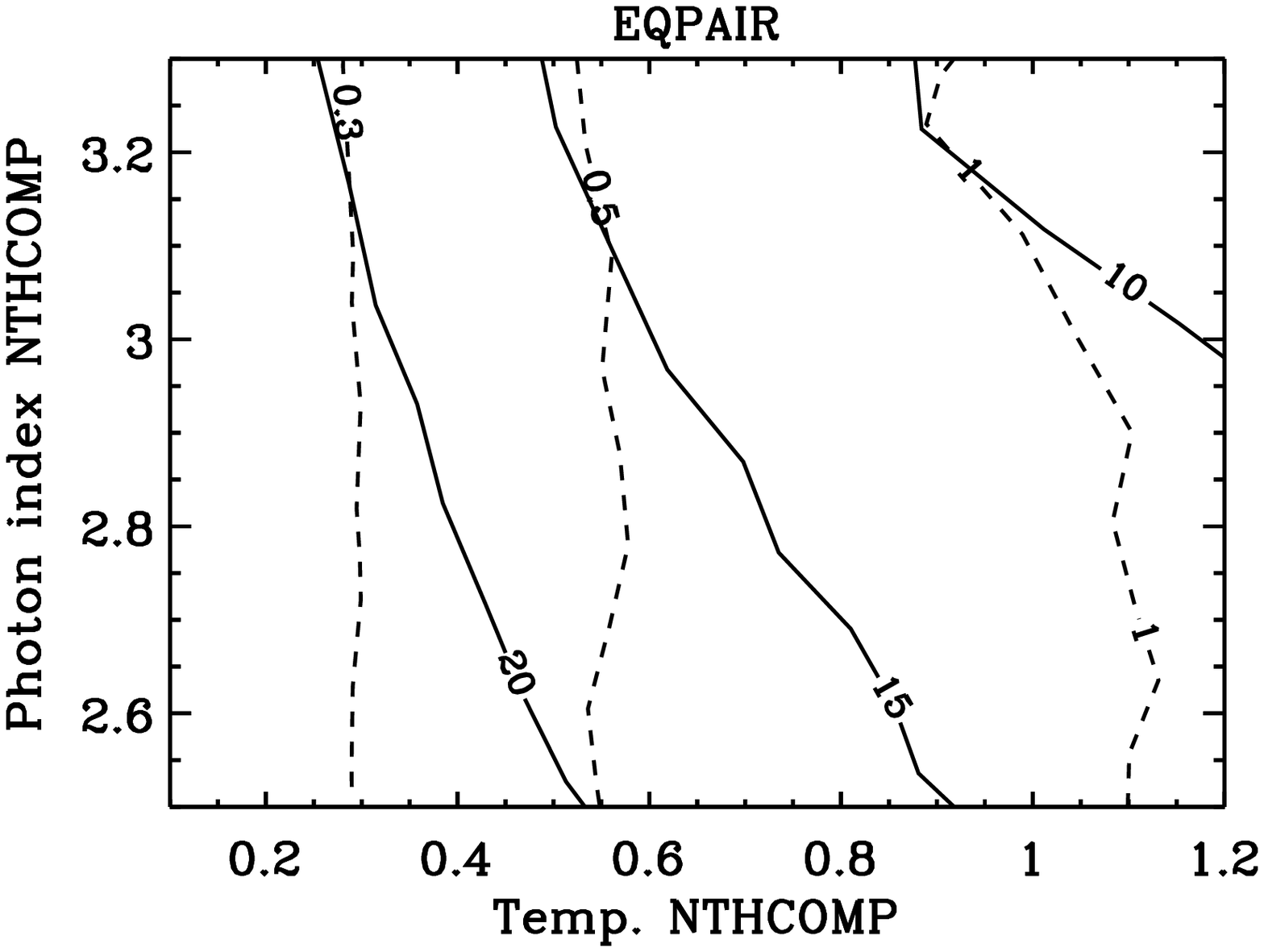}&
\includegraphics[width=\columnwidth]{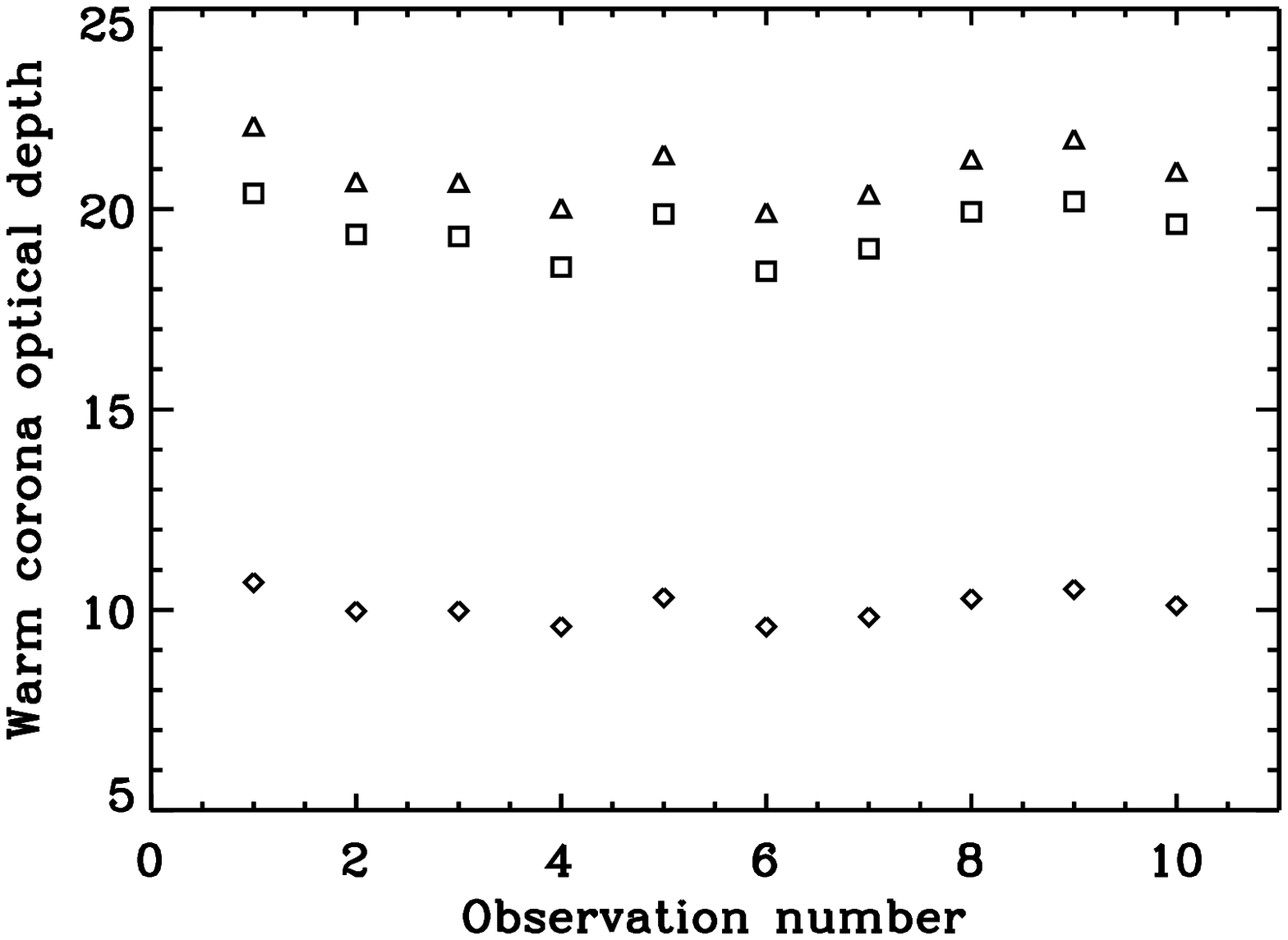}\\
\end{tabular}
 \caption{Contours of the temperature (dashed lines) and optical depth (solid lines) in the $\Gamma$-$kT_{nth}$ parameter space of \nthcomp\  for \compst\ ({\bf top left}), \comptt\ ({\bf top right}) and \eqpair\ ({\bf bottom left}) respectively. {\bf Bottom right:} Light curves of the optical depth estimates during the whole campaign for the three models: diamonds: \compst, squares: \eqpair; triangles:\comptt}
%
\label{nthvscompto}
\end{figure*}

We report in Figs. \ref{nthvscompto}a, b, and c  the contours of the temperature and optical depth for the three models \compst, \comptt, and \eqpair\ in the $\Gamma$-$kT_{nth}$ parameter space of \nthcomp. We restrict our analysis  to low temperatures and steep spectra as expected for the {\sc warm} corona. While the contours of \compst\ and \eqpair\ look relatively similar, the optical depth estimated with \comptt\ is a factor two smaller. The corresponding optical depth estimates  for the ten different observations of Mrk 509 discussed in this paper are plotted in Fig. \ref{nthvscompto}d. Possible explanations of this difference can be found in the different model assumptions, especially concerning the geometry of the disk-corona systems. It is also worth noting that the best fit temperature values of the {\sc warm} corona obtained by M11 are significantly lower (by a factor 4) than ours. In this case, however, the main reason for this difference is certainly in the use of a different model for the primary continuum (M11 use a power law). Interestingly enough, while the exact values of the optical depth clearly depend on the model used (cf. Fig. \ref{nthvscompto}d), its variability throughout the monitoring looks very similar and is apparently independent of the model used.\\

\section{Estimation of the $L_{s}$ and $L_{tot}$}
\label{ampl}

To estimate the seed soft photon luminosity entering and cooling a corona of optical depth $\tau$, we use the conservation of the number of photons, which characterizes the Compton process. From the observed comptonized spectrum, we deduce an observed photon rate $n_{obs}$. By conservation, this photon rate is equal to the sum of the seed photons crossing the corona without being comptonized $n_{s,0}$ and those comptonized in the corona and emitted upward (in direction to the observer) $n_{c,up}$. At first order,
\begin{eqnarray}
n_{s,0}&=&n_se^{-\tau}\label{eqns1}\\
n_{c,up}&=&\frac{n_s(1-e^{-\tau})}{2},\label{eqns2}
\end{eqnarray}
the factor  $1/2$ of the last equation assumes the Compton scattering process to be isotropic  with half of the comptonized photons being emitted upward, and the over half being emitted downward. The Compton process is certainly not isotropic, especially in the case of an anisotropic seed soft photon field (e.g. \citealt{haa91,haa93,ste95,hen97}), but the effect on the photon rate is relatively small.\\

Then we have 
\begin{equation}
n_{obs}=n_{s,0}+n_{c,up}=\frac{n_s(1+e^{-\tau})}{2},
\end{equation}
which gives
\begin{equation}
n_{s}=2\frac{n_{obs}}{(1+e^{-\tau})}.
\end{equation}
For an optically thin corona ($\tau\ll 1$), $n_s\simeq n_{obs}$ (very few soft photons are comptonized). This is rougly the case of the \hot\ corona. On the other hand, for $\tau\gg 1$, $n_s\simeq 2 n_{obs}$, which corresponds to the case of the \warm\ corona. Once the photon rate $n_s$ of the multicolor accretion disk is known, as is its temperature (deduced from the fit), we have access to the corresponding luminosity $L_s$.\\

The corona total power $L_{tot}$ is the sum of $L_s$ and the heating power $L_h$ liberated in the corona to comptonize the soft photon field:
\begin{equation}
L_{tot}=L_h+L_s.\label{eqtot}
\end{equation}
Here, $L_h$ and $L_s$ can be divided into upper and lower parts:
\begin{eqnarray}
L_h&=&L_{h,u}+L_{h,d}=2L_{h,u}\label{eqlhup}\\
L_s&=&L_{s,u}+L_{s,d}\\
&=&\underbrace{L_se^{-\tau}+\frac{L_s(1-e^{-\tau})}{2}}_{L_{s,u}} + \underbrace{\frac{L_s(1-e^{-\tau})}{2}}_{L_{s,d}}\label{eqlsup}
\end{eqnarray}
The first equation assumes an isotropic Compton process, and the second equation is obtained with the same reasoning as in Eq. \ref{eqns1} and \ref{eqns2}.\\

Then the observed luminosity is simply
\begin{equation}
L_{obs}=L_{h,u}+L_{s,u},\label{eqobs}
\end{equation}
where we neglect any extra emission that does not come from the corona. From Eqs. \ref{eqlhup}, \ref{eqlsup}, and \ref{eqobs}, we can deduce $L_h$
\begin{equation}
L_h=2L_{obs}-L_s(1+e^{-\tau}),
\end{equation}
and finally, from Eq. \ref{eqtot}
\begin{equation}
L_{tot}=2L_{obs}-L_se^{-\tau}.
\end{equation}
For an optically thin corona ($\tau\ll 1$), $L_{tot}\simeq 2L_{obs}-L_s$. This is roughly the case for  the \hot\ corona. On the other hand, for $\tau\gg 1$, $L_{tot}\simeq 2L_{obs}$, which corresponds to the case of the \warm\ corona.\\

The radiative equilibrium of the soft phase (which produces the soft photon field) also implies
\begin{eqnarray}
L_s &=& L_{s,d}+L_{h,d}+L_{s,intr}\\
 &=& \frac{L_s(1-e^{-\tau})}{2}+\frac{L_h}{2}+L_{s,intr}
\end{eqnarray}
where $L_{s,intr}$ is the intrinsic emission of the soft phase. In consequence, the heating/cooling ratio is equal to
\begin{equation}
\frac{L_h}{L_s}=1+e^{-\tau}-2\frac{L_{s,intr}}{L_s}.
\label{eqratio}
\end{equation} 
In the case of an optically thin ($\tau <1$) corona above a passive disk ($L_{s,intr}$=0), we found the well known result $L_{h}/L_s=2$ (see e.g. \citealt{haa91} and their Eq. 3b with f=0 and no disk albedo) and consequently $L_{tot}/L_s=3$. On the other hand, for an optically thick corona (still above a passive disk), we find $L_{h}/L_s=1$ and $L_{tot}/L_s=2$

\end{appendix}

\end{document}